\documentclass[reprint,superscriptaddress,amsmath,amssymb,aps,prb]{revtex4-2}

\usepackage{graphics,epsf}

\usepackage{graphicx}
\usepackage{dcolumn}
\usepackage{bm}
\usepackage[colorlinks=true, allcolors=blue]{hyperref}

\usepackage[capitalise]{cleveref}
\usepackage{amsmath}
\usepackage{amssymb}
\usepackage{amsfonts}

\usepackage{epstopdf}
\usepackage{amsbsy}
\usepackage[LGR,T1]{fontenc}
\usepackage{lmodern}
\usepackage{textgreek}

\usepackage{bbm}
\usepackage{braket}
\usepackage{xcolor}
\usepackage{float}
\usepackage[utf8]{inputenc}

\newcommand{\eV}{~\rm eV}
\newcommand{\meV}{~\rm meV}
\newcommand{\K}{~\rm {K}}

\newcommand{\T}{~\rm T}
\newcommand{\mT}{~\rm mT}

\begin{document}
\preprint{APS/123-QED}

\title{Ising superconductivity in the bulk incommensurate layered material $\left(\rm{PbS}\right)_{1.13} \left(\rm{TaS}_2\right)$}

\author{Sajilesh K. P.}
\thanks{These authors contributed equally}
\affiliation{Department of Physics, Technion, Haifa, 3200003, Israel}

\author{Roni Anna Gofman}
\thanks{These authors contributed equally}
\affiliation{Department of Physics, Technion, Haifa, 3200003, Israel}

\author{Yuval Nitzav}
\affiliation{Department of Physics, Technion, Haifa, 3200003, Israel}

\author{Avior Almoalem}
\affiliation{Department of Physics, Technion, Haifa, 3200003, Israel}

\author{Ilay Mangel}
\affiliation{Department of Physics, Technion, Haifa, 3200003, Israel}

\author{Toni~Shiroka}
\affiliation{PSI Center for Neutron and Muon Sciences CNM, CH-5232, Villigen PSI, Switzerland}
\affiliation{Laboratorium für Festkörperphysik, ETH Zürich, CH-8093 Zürich, Switzerland}

\author{Nicholas C. Plumb}
\affiliation{Photon Science Division, Paul Scherrer Institut, CH-5232 Villigen PSI, Switzerland}
\author{Chiara Bigi}
\affiliation{SOLEIL Synchrotron, L’Orme des Merisiers, Départementale 128, 91190 Saint-Aubin, France}

\author{Francois Bertran}
\affiliation{SOLEIL Synchrotron, L’Orme des Merisiers, Départementale 128, 91190 Saint-Aubin, France}
\author{J. Sánchez-Barriga}
\affiliation{Helmholtz-Zentrum Berlin für Materialien und Energie, Albert-Einstein-Strasse 15, 12489 Berlin, Germany}
\affiliation{ IMDEA Nanoscience, C/ Faraday 9, Campus de Cantoblanco, 28049 Madrid, Spain}
\author{Amit Kanigel}
\affiliation{Department of Physics, Technion, Haifa, 3200003, Israel}

\begin{abstract}
Exploiting the spin-valley degree of freedom of electrons in materials is a promising avenue for energy-efficient information storage and quantum computing.  A key challenge in utilizing spin-valley polarization is the realization of spin-valley locking in bulk systems. Here, we report a comprehensive study of the noncentrosymmetric bulk misfit compound \(\left(\rm{PbS}\right)_{1.13} \left(\rm{TaS}_2\right)\), showing a strong spin-valley locking. Our investigation reveals Ising superconductivity with a transition temperature of 3.14 K, closely matching that of a monolayer of TaS$_2$. Notably, the absence of charge density wave (CDW) signatures in transport measurements suggests that the PbS layers primarily act as spacers between the dichalcogenide monolayers. 

This is further supported by angle-resolved photoemission spectroscopy (ARPES), which shows negligible interlayer coupling, a lack of dispersion along the $k_{\perp}$ direction and significant charge transfer from the PbS to the TaS$_2$ layers.
Spin resolved ARPES shows strong spin-valley locking of the electronic bands.

Muon spin rotation experiments conducted in the vortex phase reveal an isotropic superconducting gap. However, the temperature dependence of the upper critical field and low-temperature specific heat measurements suggest the possibility of multigap superconductivity.

These findings underscore the potential of misfit compounds as robust platforms for both realizing and utilizing spin-valley locking in bulk materials, as well as exploring proximity effects in two-dimensional structures.

\end{abstract}

\maketitle

\section{\label{sec:Intro} INTRODUCTION}
Two-dimensional transition metal dichalcogenides (TMDCs) have garnered renewed attention in recent years owing to the emergent topological phases and unusual superconductivity \cite{TMDC,TMDC2}. In addition, leveraging the energy valleys in the band structures of TMDC materials offers a unique opportunity to harness the spin and valley degrees of freedom for valleytronics applications \cite{valley}. Apart from the structural varieties, these materials offer great flexibility in terms of tuning the electronic properties easily by different means, such as doping, intercalation, gating, and heterostructure designing\cite{tuning, intercalation,doping, gating}. Recent years have seen a surge of interest in designing heterostructured materials with electronic responses different from their constituent layers\cite{hetero,hetero2}. 
The synergy between different material platforms holds immense potential to host novel correlated phases of matter, challenging the conventional understanding and opening the door to new physics and diverse technological applications.\cite{tsc,tsc2,majorana}. However, realizing high-quality, scalable systems remains a roadblock for further experimentation. The difficulty in obtaining clean interfaces and the tedious process of device fabrication are among the major challenges in this domain \cite{issues,issues2}.

Many of these challenges related to synthesizing heterostructures can be circumvented by studying naturally grown heterostructures, such as misfit materials. Misfit materials hold a general formula of \(\left(\rm{MX}\right)_{1+x} \left(\rm{TX}_2\right)_{n}\) with stacking of monochalcogenide $\left(\rm{MX}\right)$ and dichalcogenide $\left(\rm{TX}_2\right)$ layers\cite{misfit1,misfit2}. The monochalcogenide layers crystallize into distorted NaCl-type structures, while the dichalcogenide layers consist of T atoms coordinated in a trigonal prismatic arrangement. Due to the fundamentally different symmetry and periodicity of the sublattice, the crystal structure is incommensurate along one of the in-plane crystallographic axes and remains commensurate along the other two axes \cite{misfit3}.

 Although the sub-lattices are incommensurate, causing a continuous change in the coordination number of the interface atoms, the crystals show high stability with perfect stacking perpendicular to the layers \cite{misfit2,misfit3}. Furthermore, the incommensurability prevents the formation of any strong interlayer bonding. This structure can also be considered an intercalation of monochalcogenide layers between the weakly bonded TMDC layers. The monochalcogenide spacer layer thus reduces or completely decouples the TMDC monolayers and protects them. This method enables the achievement of a 2D nature in bulk van der Waals materials, providing robust stability. Similar to the intercalation of foreign atoms or molecules, recent reports show a substantial charge transfer between the layers, which helps stabilize the complex structure\cite{misfit2,charge1,charge2}. However, the spacer layer's impact on the TMDC layer's electronic structure, and vice versa, along with the tailored electronic structure of misfit materials, has been less explored. In addition, a few misfit materials have also shown superconductivity, primarily believed to be arising from the TMDC layers \cite{misfit4,misfit5,misfit6, misfit7,charge2}. However, the nature of the superconducting gap and the superconducting pairing mechanism in these misfit compounds needs to be studied further.

Here, we have synthesized the misfit material $\left(\rm{PbS}\right)_{1.13} \left(\rm{TaS}_2\right)$, consisting of alternating layers of monochalcogenide $\rm{PbS}$ and monolayer $\rm{1H-TaS}_2$. $\left(\rm{PbS}\right)$ is a semiconductor with a direct band gap, while $\rm{2H-TaS}_2$ is a superconductor below $0.8\K$ \cite{PbS,TaS2}. The compound $\left(\rm{PbS}\right)_{1.13} \left(\rm{TaS}_2\right)$ is particularly intriguing due to its structural similarity to $\rm{4Hb-TaS}_2$, which consists of alternating layers of $\rm{1H-TaS}_2$ and strongly correlated $\rm{1T-TaS}_2$. Notably, $\rm{4Hb-TaS}_2$ has exhibited signs of topological surface superconductivity with broken time-reversal symmetry (TRS). While various measurements suggest the presence of non-trivial superconductivity in this material, the exact underlying mechanism remains unknown \cite{4hb1,4hb2,4hb3,4hb4}.

Our measurements have shown superconductivity in $\left(\rm{PbS}\right)_{1.13} \left(\rm{TaS}_2\right)$ at $3.14\K$ with a possible two-gap nature. Our transport and angle-resolved photoemission spectroscopy
(ARPES) data show the two-dimensional nature of superconductivity with strong spin-valley locking. The electronic band structure at the Fermi level is dominated by the TMDC layer, which has a large charge transfer from mono-chalcogenide to the dichalcogenide layer.

\section{\label{sec:results}RESULTS AND DISCUSSION}

\subsection{\label{subsec:crystallography} Crystallography}
The crystal structure of \(\left(\mathrm{PbS}\right)_{1.13} \left(\mathrm{TaS}_2\right)\) consists of alternating face-centered orthorhombic unit cells from each sublattice. The \(\mathrm{TaS}_2\) sublattice consists of a monolayer of noncentrosymmetric 1H-\(\mathrm{TaS}_2\), where Ta atoms are trigonal prismatically coordinated to six sulfur atoms. While \(\mathrm{PbS}\) layer adopts a distorted NaCl structure with a double layer of lead and sulfur. Unlike $\rm{4Hb-TaS}_2$ where trigonal prismatic Ta-S layers are rotated by 180$^{\circ}$ degrees relative to each other, the \(\mathrm{TaS}_2\) layers in \(\left(\mathrm{PbS}\right)_{1.13} \left(\mathrm{TaS}_2\right)\) maintain the same orientation. Furthermore, the \(\mathrm{TaS}_2\) layers are shifted by half a unit cell along the \(a\)-axis, making the crystal lattice highly noncentrosymmetric. Although the crystallographic axes of the \(\mathrm{TaS}_2\) and \(\mathrm{PbS}\) layers are parallel, the system features an incommensurate \(a\)-axis, which highlights the complex structural interplay within this misfit layer compound. Specifically, the unit cell dimensions for \(\mathrm{TaS}_2\) are \(a = 3.304 \, \text{\AA}\), \(b = 5.779 \, \text{\AA}\) and \(c = 24.002 \, \text{\AA}\). While for \(\mathrm{PbS}\), the dimensions are \(a = 5.803 \, \text{\AA}\), \(b = 5.772 \, \text{\AA}\), and \(c = 24.002 \, \text{\AA}\) \cite{PTSxrd}. 

\begin{figure*}[ht!]
\includegraphics[trim= 0cm 0cm 0cm 0cm,clip=true,width=0.99\textwidth]{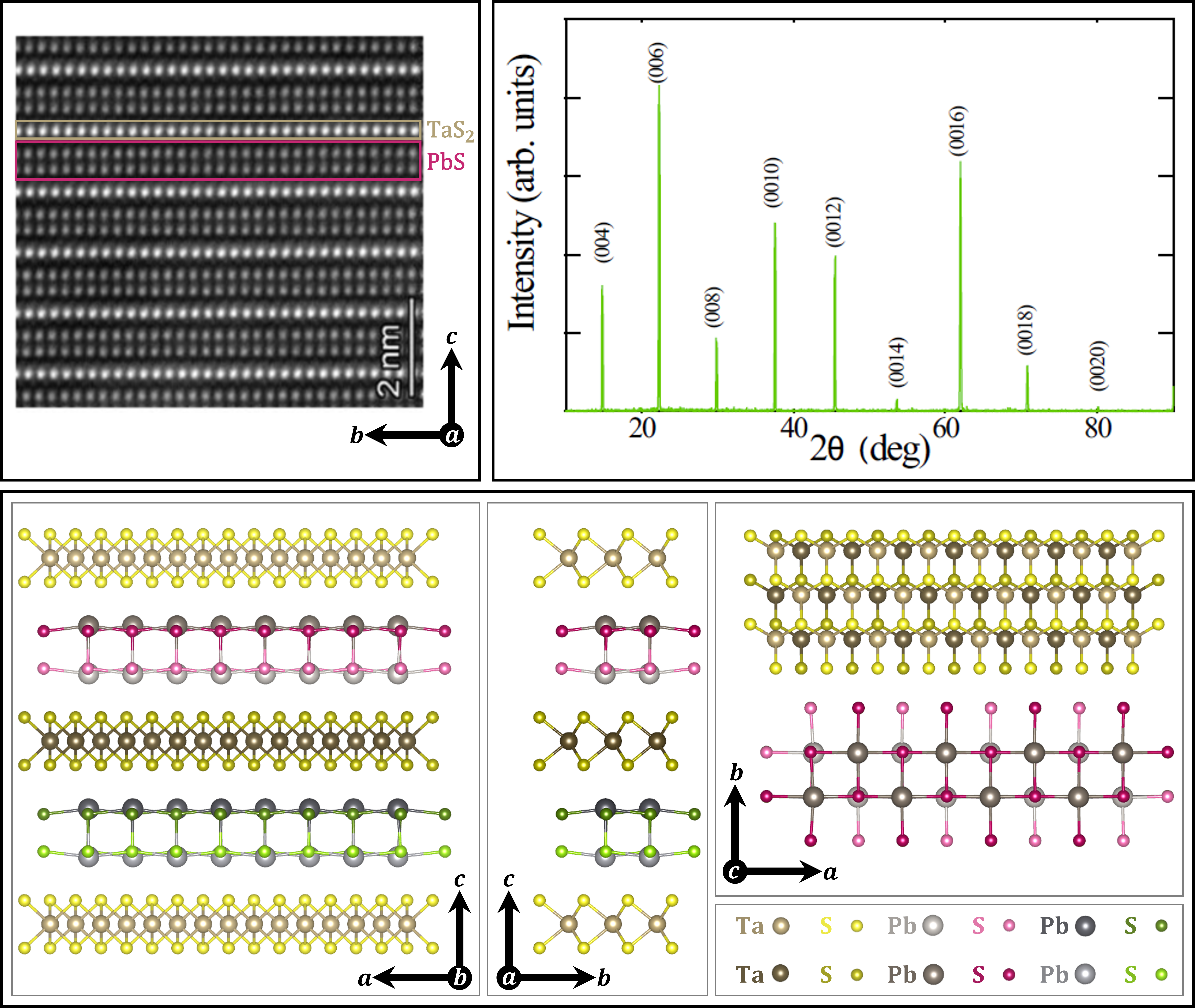}
\caption{\label{fig:PTS_StrucXRDandTEM} \textbf{Bottom:} The crystal structure of  \(\left(\rm{PbS}\right)_{1.13} \left(\rm{TaS}_2\right)\) viewed along different directions. \textbf{Top Left:} High-resolution TEM pattern showing the commensurate b-axis and perfect stacking along the c-direction \textbf{Top Right:} The single crystal x-ray diffraction pattern of the sample indicates the diffraction along the \((00l)\) plane.}
\end{figure*}

This unique structural arrangement is depicted schematically in the bottom panel of  \cref{fig:PTS_StrucXRDandTEM}. The intricate stacking of the layers is further confirmed using HR-TEM. The top left panel of \cref{fig:PTS_StrucXRDandTEM} presents a cross-sectional HR-TEM image, showcasing the well-ordered stacking of distorted NaCl-structured \(\mathrm{PbS}\) and single-layered \(\mathrm{1H-TaS}_2\) sub-lattices. The crystallographic parameters obtained from the HR-TEM match well with reported data \cite{PTSxrd}. Additionally, the XRD pattern of the as-grown single crystals, shown in the top right panel of \cref{fig:PTS_StrucXRDandTEM}, displays distinct diffraction peaks from the \((00l)\) planes, further confirming the periodicity and alignment of the layers.

.

\subsection{\label{Magnetization} Magnetization}
\begin{figure}[ht!]
\includegraphics[width=0.99\columnwidth]{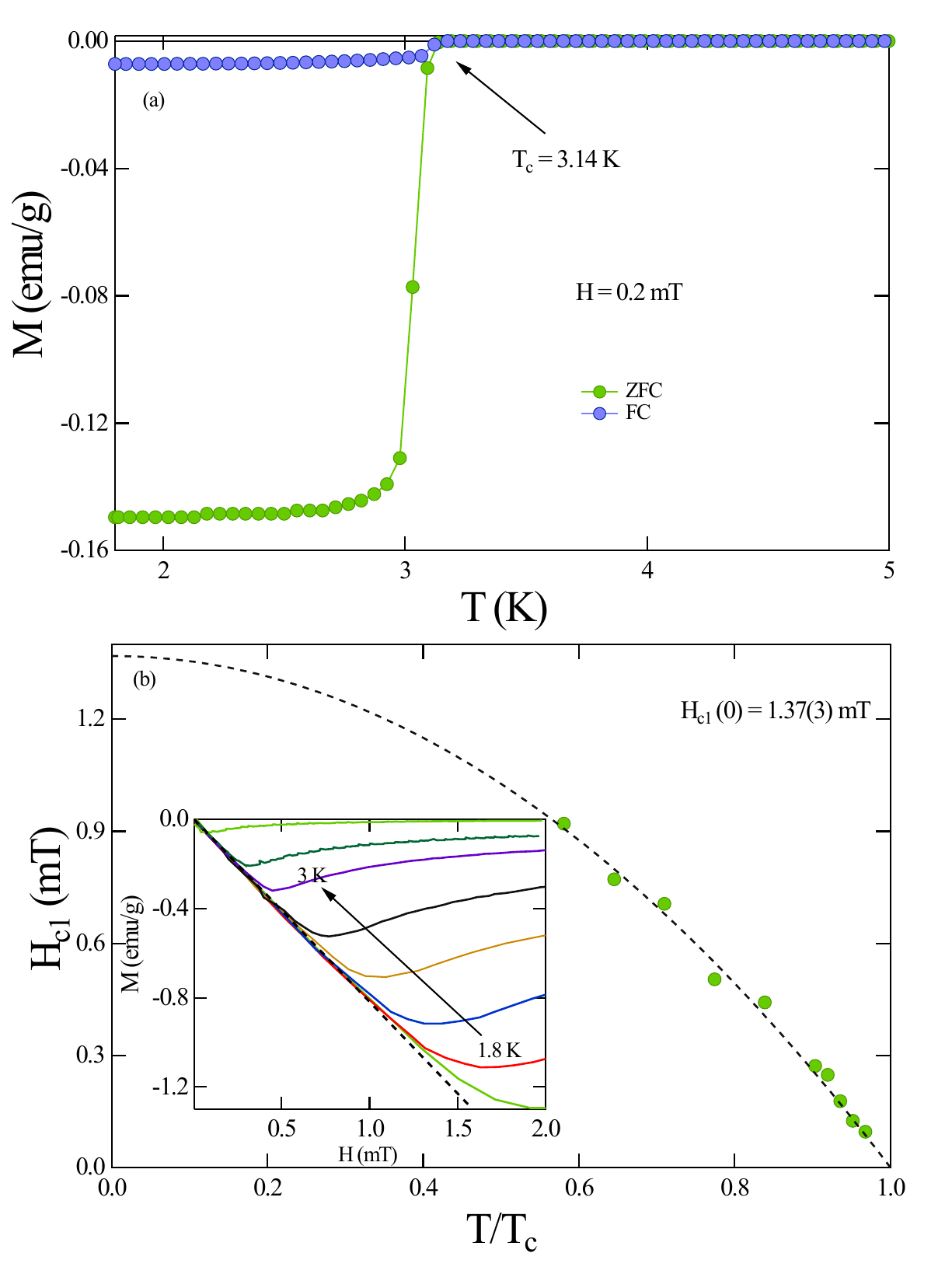}
\caption{\label{fig:PTS_tempDepend}
(a) Temperature-dependent DC magnetization data collected at \(0.2 \mT\), illustrating the onset of the superconducting transition at \(T_c = 3.14 \K\). (b) Main panel: Lower critical field \(H_{c1}\) extrapolated (black dashed line) using the G-L equation, yielding \(H_{c1} = 1.37 \mT\). Inset: Isothermal magnetization curves demonstrating the method for determining \(H_{c1}\). }
\end{figure}

The bulk superconducting nature of \(\left(\mathrm{PbS}\right)_{1.13} \left(\mathrm{TaS}_2\right)\) single crystals was confirmed through DC magnetization measurements done at 0.2 mT. Zero field cooling (ZFC) measurements revealed a diamagnetic transition at \(3.14\K\), indicative of bulk superconductivity (\cref{fig:PTS_tempDepend}a). Additionally, the type-II nature of the superconductivity was confirmed through the behavior observed in the Field cooling (FC) measurements.  The lower critical field, \(H_{c1}\), was measured as a function of temperature (\cref{fig:PTS_tempDepend}b). At each specified temperature, isothermal low-field magnetization measurements were conducted with the magnetic field applied parallel to the c-axis (\cref{fig:PTS_tempDepend}b inset). The field of deviation from linearity in the magnetization data is \(H_{c1}\) at each temperature. By employing the Ginzburg-Landau (G-L) relation \(H_{c1}(T) = H_{c1}(0)(1-(T/T_c)^2)\), we concluded \(H_{c1}(0)=1.37 \mT\). 

\subsection{\label{subsec:resistivity}Resistivity} 
 The electrical transport measurements for the \(\left(\mathrm{PbS}\right)_{1.13} \left(\mathrm{TaS}_2\right)\) sample were conducted in the \(ab\)-plane, both in zero field and under various applied magnetic fields and angles relative to the \(ab\)-plane.
 
\begin{figure*}[ht!]
\includegraphics[width=0.9\textwidth]{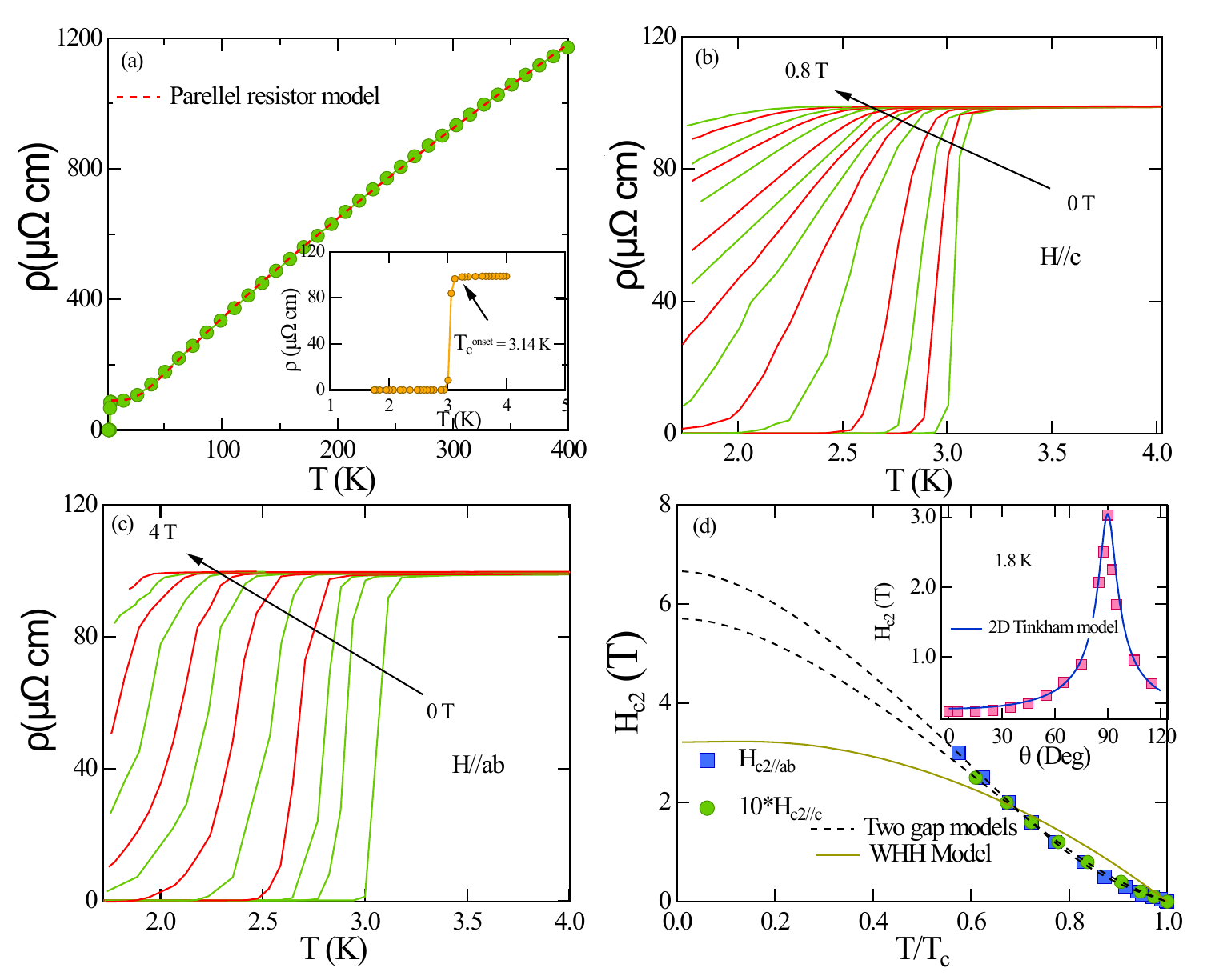}
\caption{ \label{fig:PTS_Res} 
\textbf{(a)} Electrical resistivity of \(\left(\mathrm{PbS}\right)_{1.13} \left(\mathrm{TaS}_2\right)\) over the temperature range from 400 K to 1.7 K. A dotted line represents the fit of the data using the parallel resistor model (\cref{eq:rhoT}). The inset provides a detailed view of the resistivity drop near 3.14 K, indicative of the superconducting transition.
\textbf{(b \& c)} Temperature dependence of the resistivity for \(\left(\mathrm{PbS}\right)_{1.13} \left(\mathrm{TaS}_2\right)\) measured under different applied magnetic fields, with (b) showing the out-of-plane direction and (c) the in-plane direction.
\textbf{(d)} Temperature dependence of the upper critical field for both out-of-plane and in-plane orientations. The dotted lines indicate the fits using the two-gap model, and the full line shows the fitting done using the WHH model, respectively (\cref{eq:whhb}, \cite{WHH}). \textbf{Inset:}  The angular dependence of the $H_{c2}$, fitted using the 2D-Tinkham model (\cref{eq:Hc2ThetaT})}
\end{figure*}

\cref{fig:PTS_Res}a shows the electrical resistivity at zero field in the range \(400 \K\) to \(1.8 \K\). The decrease in resistivity with decreasing temperature highlights the metallic nature of the sample, underscored by a residual resistivity ratio (RRR) of \(\rho_{300}/\rho_{10} = 10.5\), indicative of the sample’s high-quality single crystalline nature. A pronounced resistivity drop at \(3.14 \K\) (\cref{fig:PTS_Res}a inset) marks the onset of superconducting transition, with a transition width \(\Delta T = 0.16 \K\). This transition temperature \(T_c = 3.14 \K\) is considerably higher than that observed in pristine bulk \(\mathrm{TaS}_2\) (\(T_c = 0.8 \K\)) and other similar misfit layered materials. Studies have shown that the superconducting transition temperature in \(\mathrm{2H-TaS_2}\) increases as the number of layers decreases \cite{TaS2mono1}. Interestingly, monolayer \(\mathrm{TaS_2}\) exhibits a substantially higher transition around \(3.4 \K\), contrasting with \(\mathrm{NbSe}_2\), where \(T_c\) decreases with reduced thickness \cite{TaS2mono2}. 

We do not observe any signature of a CDW formation in the resistivity curve akin to the monolayer \(\mathrm{TaS}_2\), typically present at \(70 \K\) in bulk samples. The similar high \(T_c\) values, together with the absence of CDW in transport measurements, suggest that the \(\mathrm{PbS}\) layers effectively decouple the \(\mathrm{TaS}_2\) monolayers, acting as a spacer.

Panels (b) and (c) of \cref{fig:PTS_Res} show the temperature dependence of resistivity for \(\left(\mathrm{PbS}\right)_{1.13} \left(\mathrm{TaS}_2\right)\) under applied magnetic fields in the out-of-plane and in-plane \(a\) - axis directions, respectively. Utilizing the 50\% reduction criterion relative to the normal state resistivity, the values of the upper critical field, \(H_{c2}(T)\), have been determined and are presented in \cref{fig:PTS_Res}(d). We have used the two-band model to capture and extrapolate the positive curvature of data points to \(0 \K\) (dotted line in \cref{fig:PTS_Res} (d)) (see appendix for details), yielding an upper critical field of \(0.57 \T\) for the out-of-plane and \(6.7 \T\) for the in-plane directions at \(0 \K\). The anisotropy factor, H$_{c2 \parallel ab}$/ H$_{c2 \parallel c}$ is  11.7. Although this model aligns well with the experimental data, caution is advised regarding the reliability of the fit due to the complexity and number of variables involved.

For in-plane magnetic fields, where orbital effects are negligible, the upper critical field in s-wave superconductors is determined by the spin degrees of freedom. This establishes an upper limit on the critical field, known as the Pauli limit, given by H$_p$=1.86T$_c$=5.7T in our material, which is lower than the extrapolated value of 6.7 T. Along with the anisotropy factor, $\gamma$ = 11.8, this suggests the layered nature of the superconductivity and possibly, Ising superconductivity \cite{NbSe22, mos2}, as was suggested in other misfit compounds \cite{misfit7,charge2}.
 
To further probe the 2D characteristics of the material, we examined the anisotropic behavior of the upper critical field. The inset of \Cref{fig:PTS_Res}(d)  displays the angular variation of the upper critical field, where \(\theta\) is the angle between the \(c\)-axis and the applied field. A sharp increase in the upper critical field at angles approaching 90 degrees (i.e., in-plane) demonstrates the anisotropy of the material. The angular dependence of \(H_{c2}\) is effectively modeled using the Tinkham formula for a 2D superconductor:
\begin{equation}
\left|\frac{H_{c2}(\theta, T) \sin \theta}{H_{c2,\perp}}\right| + \left(\frac{H_{c2}(\theta, T) \cos \theta}{H_{c2,\parallel}}\right)^2 = 1
\label{eq:Hc2ThetaT}
\end{equation}
Here, \(H_{c2,\perp}\) and \(H_{c2,\parallel}\) represent the upper critical fields perpendicular and parallel to the \(c\)-axis, respectively. The fitting closely aligns with the experimental data, affirming the 2D nature of the material.

The G-L coherence lengths in the out-of-plane (\(\xi_{ab}\)) and in-plane (\(\xi_c\)) directions were estimated using the relations \(\xi_{ab} (0) = \left(\phi_{0}/2\pi H_{c2,\perp}(0)\right)^{1/2}\) and \(\xi_{c} (0) = \left(\phi_{0}/2\pi \xi_{ab} H_{c2,\parallel}(0)\right)\), with \(\phi_{0} = 2.07 \times 10^{-15}\) T\(\cdot\)m\(^2\). This calculation yields coherence lengths of \(\xi_{ab} (0) = 23.93\) nm and \(\xi_{c} (0) = 2.06\) nm for the in-plane and out-of-plane directions, respectively. 

\subsection{\label{Heat capacity} Heat capacity}

\begin{figure}[ht!]
\includegraphics[width=0.99\columnwidth]{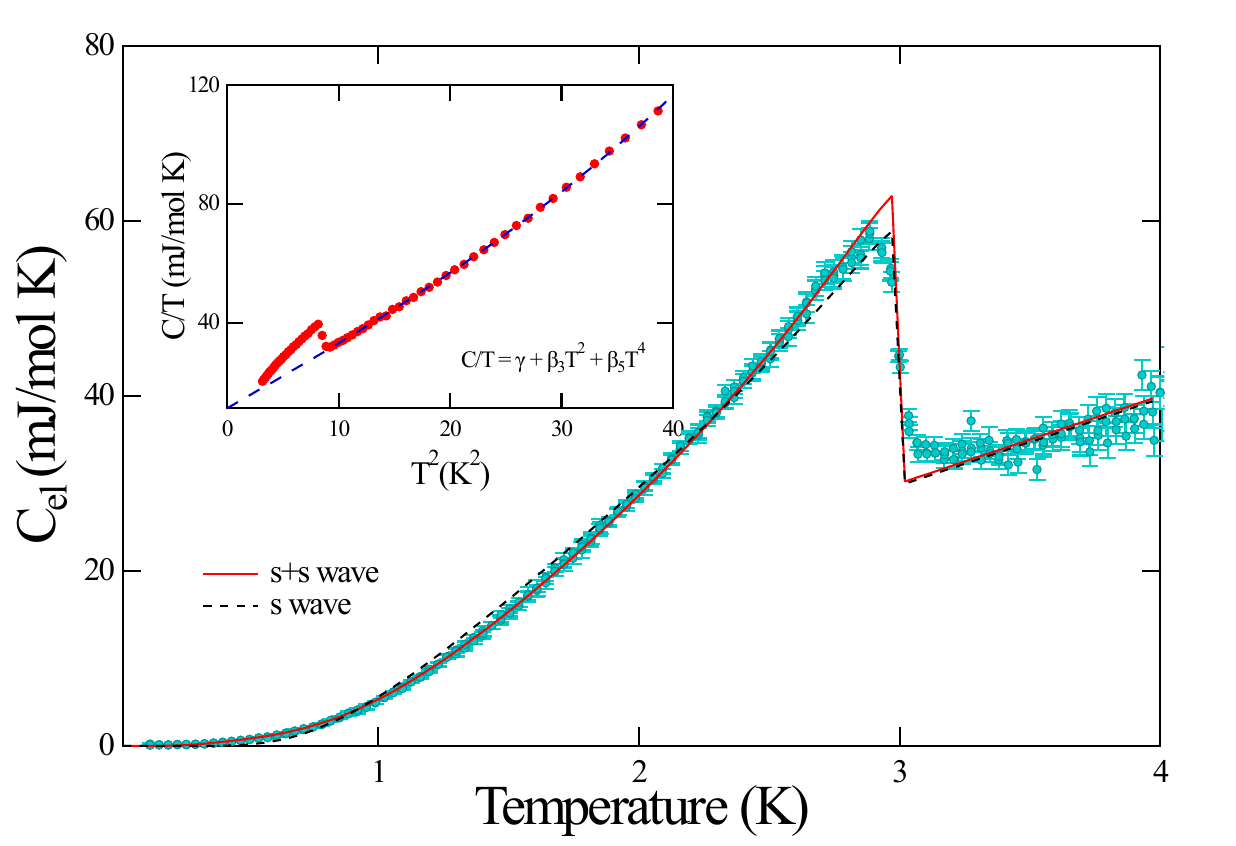}
\caption{ \label{fig:PTS_SpecHeat} Temperature dependence of the electronic specific heat in the range \(0.05 \mathrm{K} \leq T \leq 4 \mathrm{K}\) calculated by subtracting the phononic contribution. The full and dashed lines represent the fits according to the two-gap (\cref{eq:2gap}) and single-gap models (\cref{eq:Celb}), respectively.
\textbf{Inset:} The plot of \(C/T\) versus \(T^2\). The dotted line indicates the fit using \cref{eq:CtoT}.}
\end{figure}

The heat capacity measurements reveal a superconducting transition at \(T_c = 3.01 \K\) (\cref{fig:PTS_SpecHeat}). The inset in \cref{fig:PTS_SpecHeat} displays the plot of \(C/T\) versus \(T^{2}\). The data above \(T_c\) is modeled using:
\begin{equation}  
\frac{C}{T}=\gamma_{n}+\beta_{3} T^{2} + \beta_{5} T^{4}. 
\label{eq:CtoT}    
\end{equation}
here, \(\gamma_{n}\) represents the electronic contribution, while \(\beta_{3}\) and \(\beta_{5}\) are the coefficients of the phononic contribution. The fitting parameters are determined as \(\gamma_{n} = 11.72(1) \, \mathrm{mJ/mol K}^2\), \(\beta_{3} = 2.12(1) \, \mathrm{mJ/mol K}^4\), and \(\beta_{5} = 0.00029(1) \, \mathrm{mJ/mol K}^{6}\).

Using the value of \(\beta_{3}\), along with the universal gas constant \(R\) and considering 5 atoms per unit cell—while omitting the misfit index—the Debye temperature (\(\theta_{D}\)) is calculated as follows \cite{Tinkham}:
\begin{equation} 
\theta_{D} = \left(\frac{12 \pi^4 R N}{5 \beta_{3}}\right)^{\frac{1}{3}}
\label{eq:thetaD}  
\end{equation}
This equation yields a Debye temperature of \(166 \K\), which is in good agreement with the value derived from the parallel resistor model (see appendix).

The specific heat jump at the superconducting transition, given by \(\frac{\Delta C_{el}}{\gamma_{n}T_{C}} = 0.85 \pm 0.06\), is notably lower than the expected BCS value of \(1.43\). This deviation suggests the presence of non-BCS superconducting gap features, such as multigap superconductivity or gap anisotropy \cite{MgB2, anis}. 
In \cref{fig:PTS_SpecHeat} we show the electronic part of the specific heat, \(C_{el}\), extracted from the data by subtracting the phononic contribution.  The observed exponential increase in \(C_{el}\) at low temperatures is characteristic of a full superconducting gap with no nodes \cite{BCS}.

For a conventional BCS system, the entropy \(S\) below \(T_c\) is described by:
\begin{equation}
\begin{aligned}
   \frac{S}{\gamma T_{c}} = &-\frac{6}{\pi^2}\left(\frac{\Delta_{0}}{k_{B}T_{c}}\right) \\
    &\times \int_0^{\infty} \left[ f \ln(f) + (1-f) \ln(1-f) \right] dy,
\end{aligned}
\label{eq:swave}
\end{equation}
where \(f(\xi) = \left[\exp(E(\xi)/k_{B}T) + 1\right]^{-1}\) is the Fermi–Dirac distribution function, \(E(\xi) = \sqrt{\xi^{2} + \Delta^{2}(t)}\) represents the quasiparticle energy, \(\xi\) is the energy measured relative to the Fermi energy, \(\Delta(t) = \tanh\left[1.82(1.018((1/t)-1))^{0.51}\right]\) models the BCS temperature dependence of the energy gap, \(\textit{y} =\xi/\Delta(0)\) and \(t = T/T_c\). Differentiating \(S\) provides the total electronic specific heat below \(T_c\) as:
\begin{equation}
\frac{C_{el}}{\gamma T_{c}} = t \frac{d(S/\gamma T_{c})}{dt}.
\label{eq:Celb}
\end{equation}
whereas  above \(T_c\), it is equal to \(\gamma T\). For the fitting, we have adopted the principles of the \(\alpha\) model, where the quantity \(\alpha \equiv \frac{\Delta_{0}}{k_{B}T_{c}}\) is treated as a variable \cite{alpha}. Given the two-gap feature revealed by the upper critical field, we examined both single-gap and two-gap models for the electronic-specific heat. In two-gap superconductors, interband coupling ensures that both gaps open at the same critical temperature \(T_c\), with the larger gap dominating the contributions at \(T_c\). The influence of the smaller gap becomes significant at lower temperatures, affecting \(C_{el}\) in this region.
The entropy \(S\) for a two-gap superconductor is modeled as:
\begin{equation}
S(T) = w S(\Delta_{0}^{1}, T) + (1-w) S(\Delta_{0}^{2}, T),
\label{eq:2gap}
\end{equation}
where \(\Delta_{0}^{1}\) and \(\Delta_{0}^{2}\) represent the two isotropic superconducting gaps with relative weight \(w\).

The data and fits shown in \cref{fig:PTS_SpecHeat} indicate that the single-band model yields a normalized superconducting gap of \(\frac{\Delta_{0}}{k_{B}T_{c}} = 1.45(4)\) with and \(\gamma = 9.92(4) \mathrm{mJ/mol K}^2\), corresponding to \(\Delta_{0} = 0.38(4) \meV\). While more accurate fitting is obtained using the two-band model which estimates the superconducting gap values as \(\frac{\Delta^{1}_{0}}{k_{B}T_{c}} = 1.68(2)\) and \(\frac{\Delta^{2}_{0}}{k_{B}T_{c}} = 0.53(3)\) with a weighting factor of \(w=0.83\), and \(\gamma = 10.01(4) \mathrm{mJ/mol K}^2\). These values corresponds to gap values of \(\Delta_{0} = 0.44(2)\) meV  and \(0.13(2)\) meV, respectively. During the fitting procedure, the superconducting transition temperature was fixed to \(T_c = 3.01 \K\) for both models.

\subsection{\label{Muon spin rotation} Muon spin rotation}
To further investigate the temperature dependence and symmetry of the superconducting gap, we conducted transverse field muon spin rotation (TF-$\mu$SR) measurements within the vortex state of the superconductor. In the vortex state, the broadening of magnetic field distribution, due to the flux line lattice (FLL) arrangement, was evident from changes in the muon depolarization rate. The FLL distribution is closely linked to the penetration depth and, consequently, to the superconducting gap's magnitude and symmetry.

\begin{figure}[ht!]
\includegraphics[width=0.99\columnwidth]{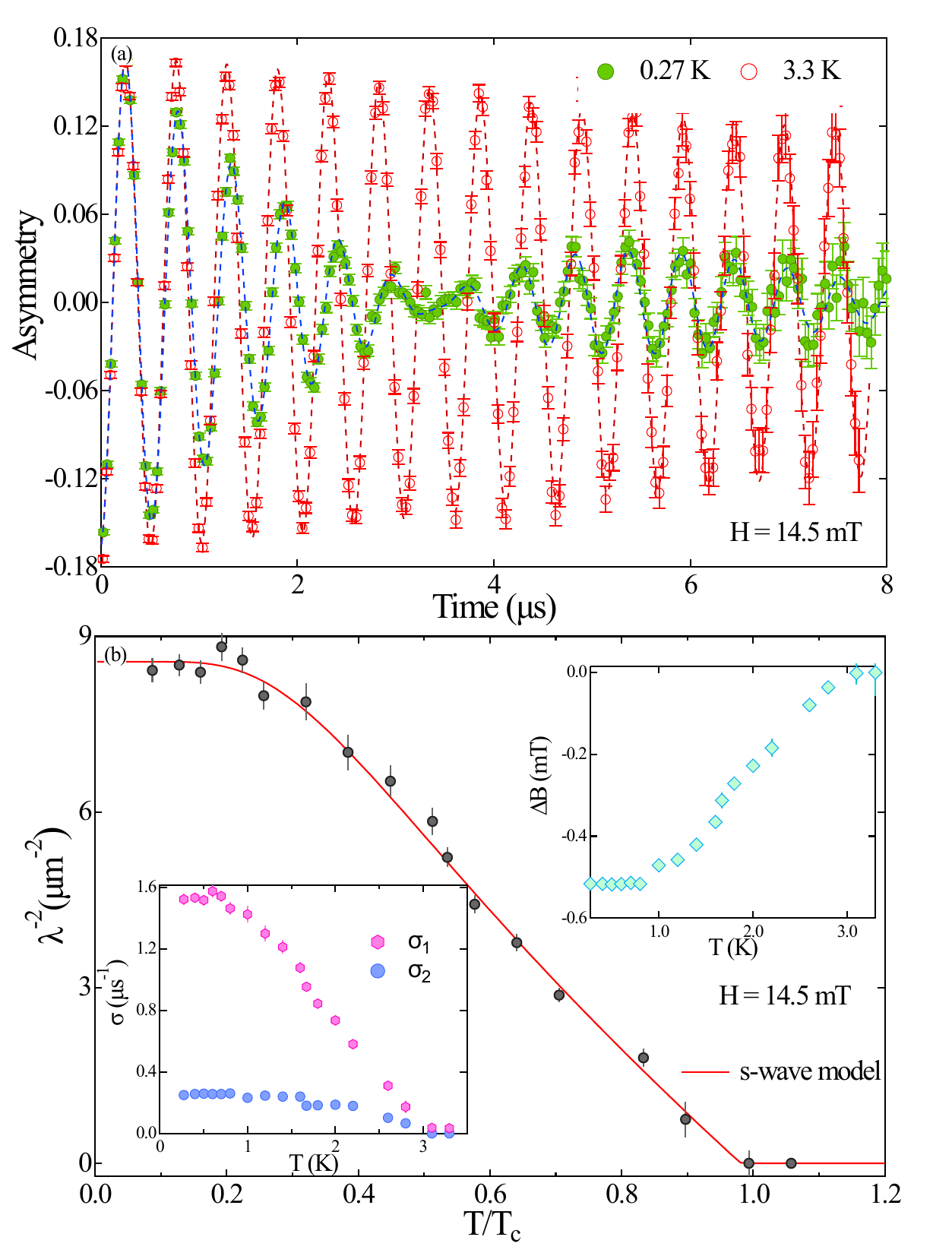}
\caption{ \label{fig:PTS_sigma} \textbf{(a)} The asymmetry spectra collected at two different temperatures across the superconducting transition. The dotted line represents the fitting done using \cref{eq:asy}. \textbf{(b)} The temperature dependence of the penetration depth. Data points are effectively modeled using the BCS single-gap (solid line) model. The left inset in (b) displays the muon depolarization rates, while the right inset presents the distribution of internal magnetic fields. Note the clear diamagnetic shift below \(T_c\)}
\end{figure}

To ensure the formation of a well-ordered FLL, the sample was field-cooled at a field of \(14.5 \mT\). Panel (a) of \cref{fig:PTS_sigma} presents typical data collected below and above $T_c$. Below $T_c$, a rapid decay in the asymmetry spectra indicates the emergence of a field distribution due to the formation of the FLL. Above $T_c$, we observe a minor decay in asymmetry, which is attributed to the relatively high nuclear magnetic moments from Ta nuclei, which remains constant across the superconducting transition.

For a comprehensive analysis, we modeled the muon asymmetry as a function of time as follows:
\begin{equation}
\begin{aligned}
        G_{\mathrm{TF}}(t) = & \sum_{i=1}^{n} A_{i}\exp\left(-\frac{\sigma^{2}_{i}t^{2}}{2}\right)\cos(\gamma_\mu B_{i}t+\phi) \\ & + A_{bg}\exp\left(-\frac{\sigma^{2}_{bg}t^{2}}{2}\right)\cos(\gamma_\mu B_{bg}t+\phi).
\end{aligned}
\label{eq:asy}
\end{equation}
Here, the first term represents the signal from the sample, and the second term pertains to background signals or signals from non-superconducting parts of the sample. \(A_i\) is the initial asymmetry, \(\sigma_i\) is the Gaussian relaxation rate, and \(B_i\) is the mean field of the \(i\)th component of the field distribution. \(\phi\) is the common phase offset, and \(\gamma_\mu/2\pi = 135.5 ~\rm{MHz/T}\) is the muon gyromagnetic ratio. The model allows \(n\) components.

 In strong type-II superconductors at high magnetic fields, the vortex density is very high, resulting in overlapping vortices and a narrow field distribution. The muon depolarization in such a state can be well explained by a single Gaussian oscillatory term.  Conversely, superconductors at low applied fields have broader field distribution, requiring multiple components to describe the asymmetry spectra accurately. In the case of multiple components, the first and second moment of the magnetic field distribution is given by \cite{secondmoment}:
\begin{equation}
\langle B \rangle = \sum_{i=1}^{n} \frac{A_i B_i}{A_{tot}},
\label{eq:firstmoment}
\end{equation}

\begin{equation}
\langle B^2 \rangle =  \frac{\sigma^2}{\gamma^2_\mu} = \sum_{i=1}^{n} \frac{A_i}{A_{tot}}\left[\frac{\sigma_i^2}{\gamma_\mu^2}-(B_i-\langle B \rangle)^2\right],
\label{eq:secondmoment}
\end{equation}
where \(A_{tot} = \sum_{i=1}^{n}A_i\). In our case, $i$ = 2 describes the asymmetry spectra very well, and it is shown as the dotted line in \cref{fig:PTS_sigma} (a). The left inset in \cref{fig:PTS_sigma} (b) shows the individual relaxation rates and the right inset shows the temperature dependence of the net change in the internal field inside the sample.  

To isolate the superconducting contribution, the relaxation rate \(\sigma_{sc}\) is derived by subtracting the square of the temperature-independent nuclear relaxation rate \(\sigma_{N}\) from the total relaxation rate \(\sigma\) as follows: \(\sigma_{sc} = \sqrt{\sigma^{2} - \sigma_{\mathrm{N}}^{2}}\).

For small applied magnetic fields, where the field \(H\) is much less than the  $H_{c2}$ (\(H / H_{c2} \ll 1\)), the penetration depth \(\lambda(T)\) can be calculated using Brandt's formula adapted for an Abrikosov vortex lattice \cite{penetrationdepth}:
\begin{equation}
    \sigma_{\mathrm{sc}}(T) = \frac{0.0609 \times \gamma_{\mu} \phi_{0}}{\lambda^{2}(T)}
    \label{eq:lamda}
\end{equation}
where \(\sigma_{\mathrm{sc}}(T)\) is expressed in \(\mu\)s\(^{-1}\), \(\lambda(T)\) is in nm, and \(\phi_{0} = 2.067 \times 10^{-15} \, \text{Wb}\) is the magnetic flux quantum.  $\lambda^{-2}(T)$ extracted using \cref{eq:lamda} is plotted in \cref{fig:PTS_sigma} (b).

The temperature dependence of \(\lambda(T)\) in the clean limit can be expressed as:
\begin{equation}
     \lambda^{-2}(T) = 1 + 2 \left\langle \int_{|\Delta(T)|}^{\infty} \left(\frac{\delta f}{\delta E}\right) \frac{E \, dE}{\sqrt{E^{2} - \Delta^{2}(T)}} \right\rangle,
\end{equation}
where \(f = [1 + \exp(E/k_{B}T)]^{-1}\) is the Fermi–Dirac distribution function. Here, the \(\Delta(T, \hat{k})\) is the temperature dependence of the superconducting energy gap. Within the BCS theoretical framework, \(\Delta(T, \hat{k})\) is modeled by:
\begin{equation}
\Delta (T) = \Delta(0) \tanh \left[1.82 \left(1.018 \left(\frac{T_c}{T} - 1\right)\right)^{0.51}\right],
\end{equation}
where \(\Delta(0)\) is the gap magnitude at zero temperature.

\cref{fig:PTS_sigma}(b) shows the fits to the data using the $s$-wave model. The estimated superconducting gap values from the $s$-wave model is \(0.35(1) \meV\), in agreement with results from the single gap heat capacity measurements. Additionally, the zero temperature penetration depth is estimated at \(\lambda (0) = 341(2) \, \text{nm}\).

We have also conducted zero-field muon spin relaxation (\( \mu \)SR) measurements at various temperatures across the superconducting transition to search for spontaneous magnetic field generation in the superconducting state, which would indicate broken time-reversal symmetry (TRS). However, our results showed no change in the asymmetry spectra above and below \( T_c \), suggesting preserved TRS in \((\text{PbS})_{1.13}(\text{TaS}_2)\). This contrasts with observations in the structurally similar \(\text{4Hb-TaS}_2\), where broken TRS was detected, implying that the 1T-TaS\(_2\) layer in \(\text{4Hb-TaS}_2\) may play a key role in its unconventional superconducting properties.   
\subsection{\label{subsec:ARPES}ARPES}
Bulk PbS is a semiconductor with a band gap of  0.2 \(eV\), but DFT calculations show that in \(\left(\mathrm{PbS}\right)_{1.13} \left(\mathrm{TaS}_2\right)\) the PbS layers should contribute to the density of state at the Fermi-level \cite{PTS_DFT}. 

\begin{figure}[ht!]
\includegraphics[width=1.0\columnwidth]{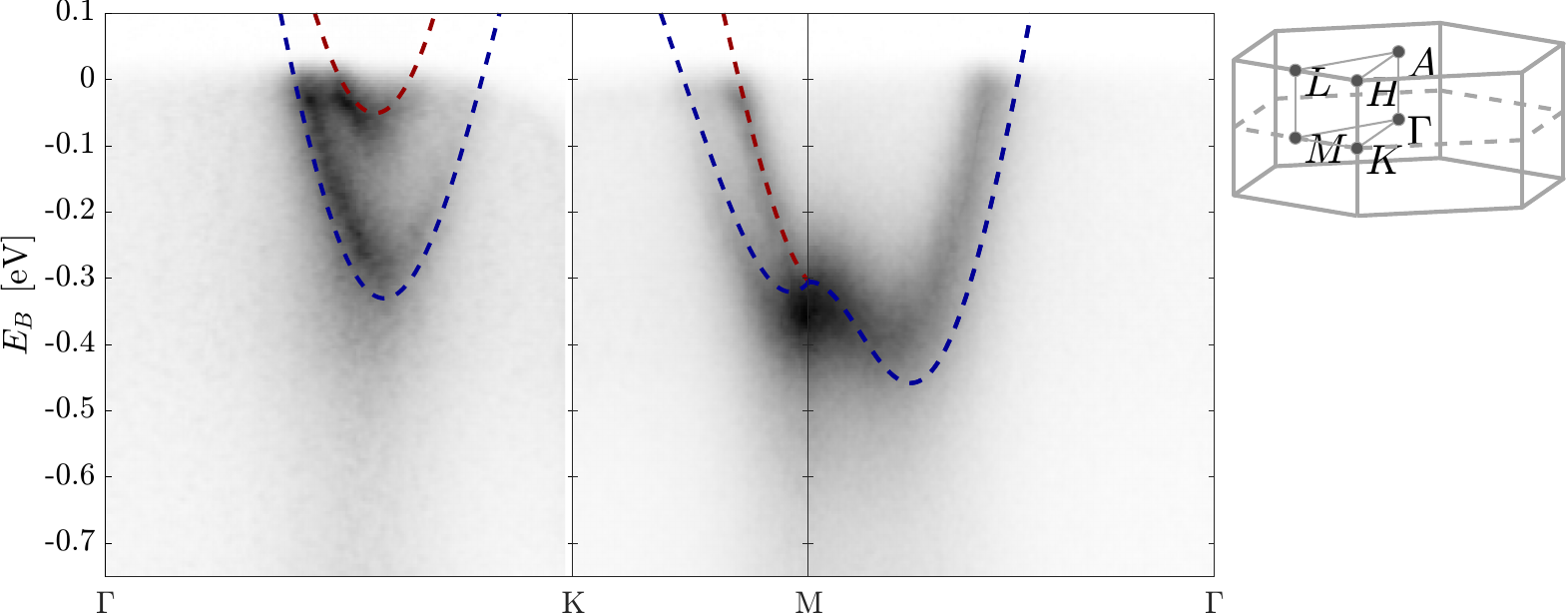}
\caption{ Measured valence band of \(\left(\mathrm{PbS}\right)_{1.13} \left(\mathrm{TaS}_2\right)\) along the high-symmetry directions \(\Gamma\mathrm{K}\), \(\mathrm{KM}\), and \(\mathrm{M}\Gamma\)  using photon energy of 100\eV. The dashed dark red and blue lines represent spin-up and spin-down valence bands, respectively, as obtained from TB calculations. \textbf{Top-Right:} Brillouin zone of the hexagonal lattice. }
\label{fig:PTS_Disp}
\end{figure}

 \Cref{fig:PTS_Disp} shows the ARPES spectra  of $\left(\mathrm{PbS}\right)_{1.13}\left(\mathrm{TaS}_2\right)$ along  the high symmetry directions $\Gamma$K, KM, and M$\Gamma$.  All of the bands crossing the Fermi level originate from the TaS$_2$ layers. Core-level spectra taken from the same location confirm the presence of both TaS$_2$ and PbS surface terminations (see \cref{fig:corelevel}). This suggests that the PbS layers may retain their semiconducting character in the PTS structure. Alternatively, the inherent disorder in the PbS layers might cause diffusive spectral features, preventing us from detecting their contribution. It is common in transition metal dichalcogenide misfit structures for the TMDC layer to dominate the ARPES spectra \cite{PbSeNbSe2, PbSeTiSe2, SnSTaS2, LaSeNbSe}. However, using nano-ARPES, the contribution of the monochalcogenide layer to the band structure was observed in (BiSe)$_{1+\delta}$NbSe$_2$ \cite{BiSe_NbSe2_ARPES}.

For the analysis of the ARPES data we assume  \(\left(\mathrm{PbS}\right)_{1.13}\left(\mathrm{TaS}_2\right)\) can be treated as monolayers of   \(1H-\mathrm{TaS_2}\) with  \(\mathrm{PbS}\) layers acting as spacers. We can describe the dispersion around the M point along the K-M-K line with  a simple model given by:
\begin{equation}
H = \frac{\hbar^2k^2}{2m^*} \sigma_0 + \alpha_\mathrm{SO} k \sigma_z 
   \label{eq:ToyHamiltonian}
\end{equation}
where \(m_*\) is the effective mass, \(\alpha_\mathrm{SO}\) is the SO coupling parameter and \(\sigma\) are Pauli matrices that represent spin direction.

\begin{figure}[ht!]
\includegraphics[width=1.0\columnwidth]{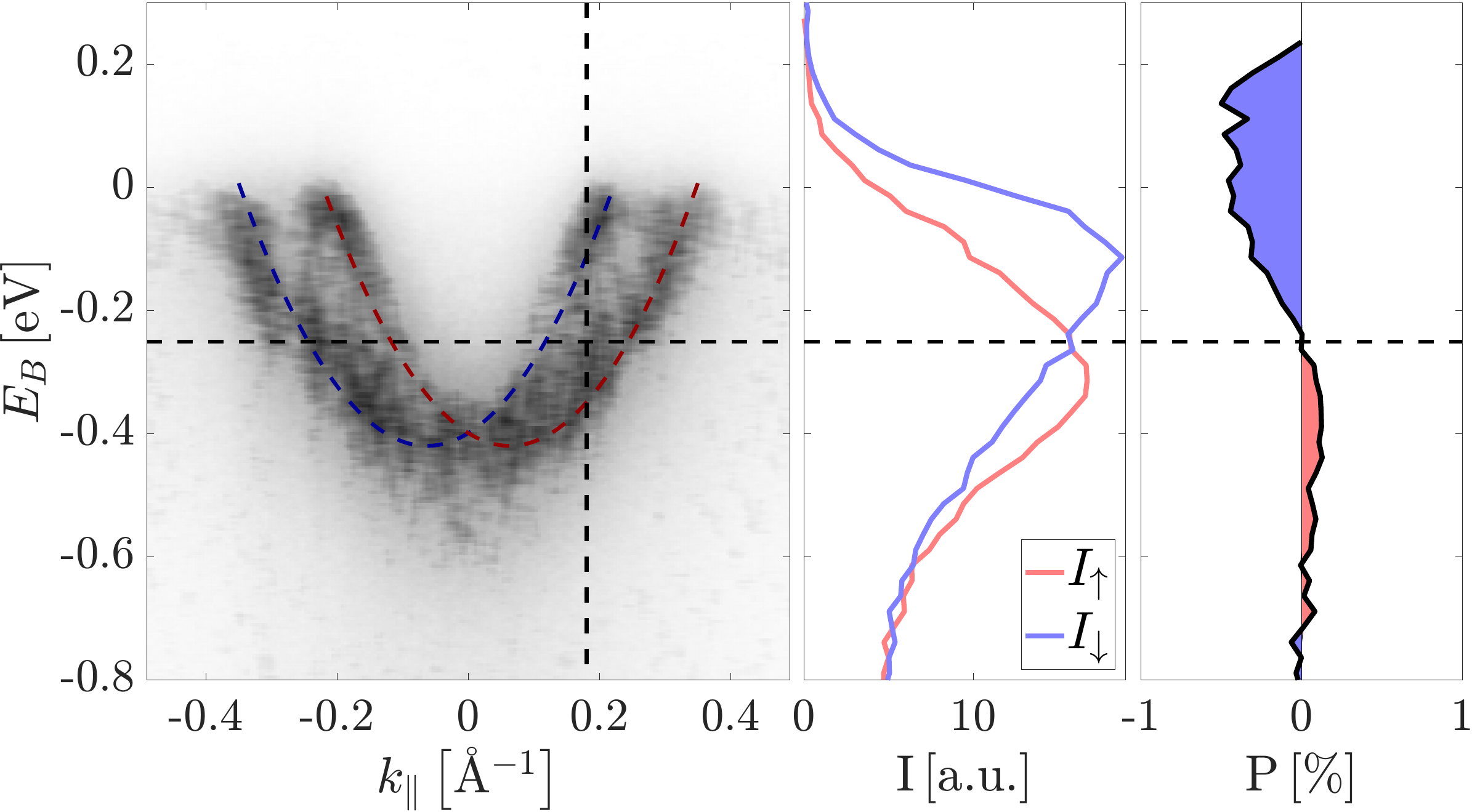}
\caption{\textbf{Left Panel:}   \(\left(\rm{PbS}\right)_{1.13} \left(\rm{TaS}_2\right)\) ARPES data along the K-M-K line measured with a photon energy of $55 \eV$. The dark red and blue lines are a fit the approximated Hamiltonian (\cref{eq:ToyHamiltonian}). The vertical dashed line marks the constant momentum line at which spin-resolved ARPES was measured. \textbf{Middle Panel:} In red (blue) is the of out-of-plane up (down) spin EDC.
\textbf{Right Panel:} The out-of-plane polarization. The horizontal dashed line marks the binding energy where the polarization switches sign.}
\label{fig:PTS_SRandDis}
\end{figure}

The left panel of \cref{fig:PTS_SRandDis} shows the ARPES spectra measured along K-M-K line at 55 \(eV\) photon energy. The red and blue dashed lines represent a fit to the data using  (\cref{eq:ToyHamiltonian}). We find an effective mass: \(m^* = 0.73 \pm 0.02 m_e\) and a SO coupling coefficient of \(\alpha_\mathrm{SO} = 657 \pm 20 \meV \cdot \AA\). The SO coupling coefficient corresponds to an average electric field strength of \(\sim 176 \; \mathrm{V}/\AA\) and matches, within the error limits the SO coupling coefficient of \(\rm{TaS_2}\) \cite{4hb4}.

The absence of inversion symmetry in the unit cell leads to SO split bands that are expected to exhibit spin polarization \cite{zhang2014hidden}. (\cref{eq:ToyHamiltonian}) produces eigenstates corresponding to pure spin-up and spin-down states. As a result, the dark red (blue) curve in the left panel of \cref{fig:PTS_SRandDis} represents spin-up (down) polarization. 

We emphasize that unlike in 2H-structured TMDs, where spin-valley locking in the two layers of the unit cell is opposite, the spin-valley locking in this system is global due to its non-centrosymmetric structure.

The middle panel of \cref{fig:PTS_SRandDis} shows the spin-resolved energy distribution curves (EDC)  measured at a fixed momentum of approximately \(0.2 \, \text{Å}^{-1}\) from the M point toward the K point, as indicated by the dashed vertical line in the left panel.
 
In the right panel, we show the spin polarization at this specific momentum, calculated as \(P = \frac{I_\uparrow - I_\downarrow}{S(I_\uparrow + I_\downarrow)}\) \cite{bawden2016spin}, where \(S = 0.16\) is a constant Sherman function. The horizontal dashed line marks the binding energy threshold where the polarization shifts from negative to positive, occurring at approximately \(-0.25 \, \eV\), coinciding with the energy transition from the upper to the lower band at the measured momentum point.

\begin{figure*}
\includegraphics[width=0.7\textwidth]{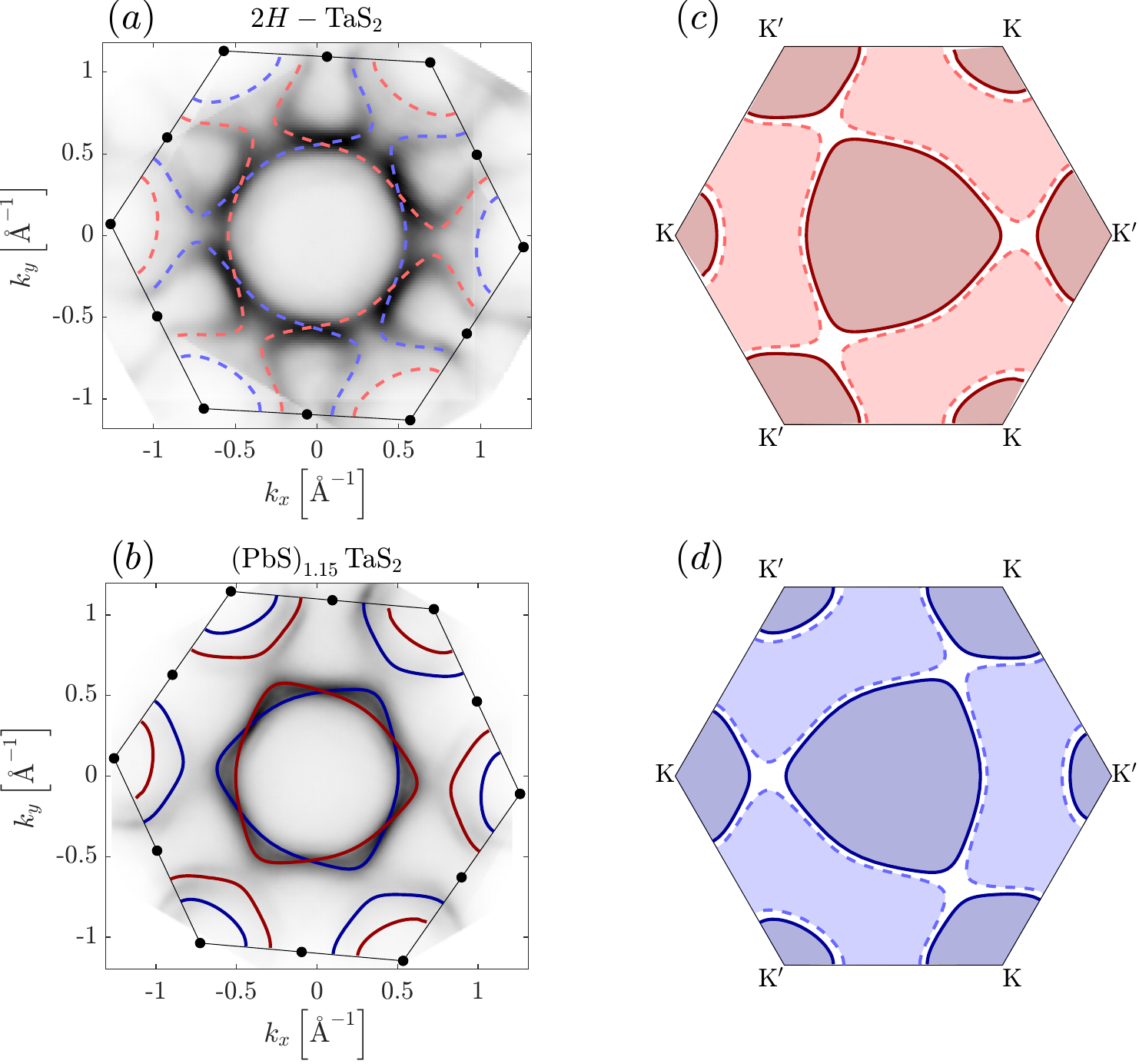}
\caption{ \textbf{(a)} Symmetrized measured Fermi surface map of \(\rm{2H-TaS}_2\). Black lines outline the 2D BZ, and black dots denote high symmetry points. The red and blue contours represent the TB calculated Fermi surface for spin-up and spin-down states, respectively. \textbf{(b)} Symmetrized measured Fermi surface map of \(\left(\rm{PbS}\right)_{1.13}\rm{TaS}_2\), annotated similarly to (a), with dark red and dark blue lines indicating energy contours at \(150\ \text{meV}\) higher than those in (a). \textbf{(c, d)} Comparison of the calculated spin-up and spin-down Fermi surfaces for \(\left(\rm{PbS}\right)_{1.13}\rm{TaS}_2\) (dark shade) and \(\rm{TaS}_2\) (light shade).}

\label{fig:CompTs2PTS}
\end{figure*}

Panels a and b of \cref{fig:CompTs2PTS} display symmetrized ARPES intensity maps at the Fermi level for  \(2H-\mathrm{TaS}_2\) and \(\left(\rm{PbS}\right)_{1.13} \left(\rm{TaS}_2\right)\), respectively. The strong resemblance between these Fermi surfaces suggests that the Fermi surface primarily arises from the \(\mathrm{TaS}_2\) bands. To fit the data, we used a tight-binding (TB) model for a single layer of \(1H-\mathrm{TaS}_2\). This model, based on ref. \cite{TB1HTaS2}, incorporates the three Ta d-orbitals (\(d_{z^2}, d_{xy}, d_{x^2-y^2}\)) and spin-orbit coupling, with \(\lambda = 0.272\) selected to align with our spin-orbit coupling parameter \(\alpha_\mathrm{SO}\). The model shows excellent agreement with the measured dispersion along the \(\Gamma\)-K and K-M directions, with only a slight deviation observed along the M-\(\Gamma\) direction, as illustrated in Fig. \ref{fig:PTS_Disp}. 
 The lines in \cref{fig:CompTs2PTS}(a and b) represent the calculated Fermi surfaces.  The chemical potential for \(\left(\rm{PbS}\right)_{1.13} \left(\rm{TaS}_2\right)\) is \(150(5) \meV\) larger than in \(2H-\mathrm{TaS}_2\) suggesting a significant charge transfer between the PbS and TaS$_2$ layers. 

In \(\rm{2H-TaS}_2\), the spin-up band forms three electron pockets around the K points, highlighted by the light-shaded regions in panel c. As electrons are transferred from the PbS layers, the system undergoes a Lifshitz transition and the Fermi-surface of \(\left(\rm{PbS}\right)_{1.13} \left(\rm{TaS}_2\right)\) consists of three-hole pockets centered around the \(\Gamma\), \(\rm{K'}\), and \(\rm{K}\) points, shown by the dark-shaded areas. The same behavior is observed for the spin-down band, but with a 60-degree rotation, as illustrated in panel d.
The difference in the Fermi-surface area indicates that  \(0.2 \pm 0.05\) electrons per unit cell are transferred from the \(\mathrm{PbS}\) layers to the \(\mathrm{TaS}_2\) layers.  

To verify the 2D nature of the electronic structure of \(\left(\rm{PbS}\right)_{1.13} \left(\rm{TaS}_2\right)\), we conducted detailed measurements of the electron dispersion normal to the TaS$_2$ planes,  along the M-L-M. To scan the out-of-plane momentum we vary the photon energy between  \(45\eV\) to \(59\eV\) covering about 2 Brillouin zones. 
We used an inner potential of \(V_0=9\eV\) \cite{4hb4}. For more details see \cref{Appendix:KZ}.  
The k$_z$ dispersion is shown in \cref{fig:kzDep}. 
Remarkably, the peak in the EDC does not change across different photon energies, showing no dispersion perpendicular to the surface. This is very different from the bulk 2H-TaS$_2$ case, where we found significant k$_z$ dispersion \cite{4hb4}.

 The lack of k$_z$ dispersion is consistent with our previous assumption that the interlayer (IL) coupling between the \(\mathrm{TaS}_2\) layers is negligible. 
 
\begin{figure}[ht!]
\includegraphics[width=0.7\columnwidth]{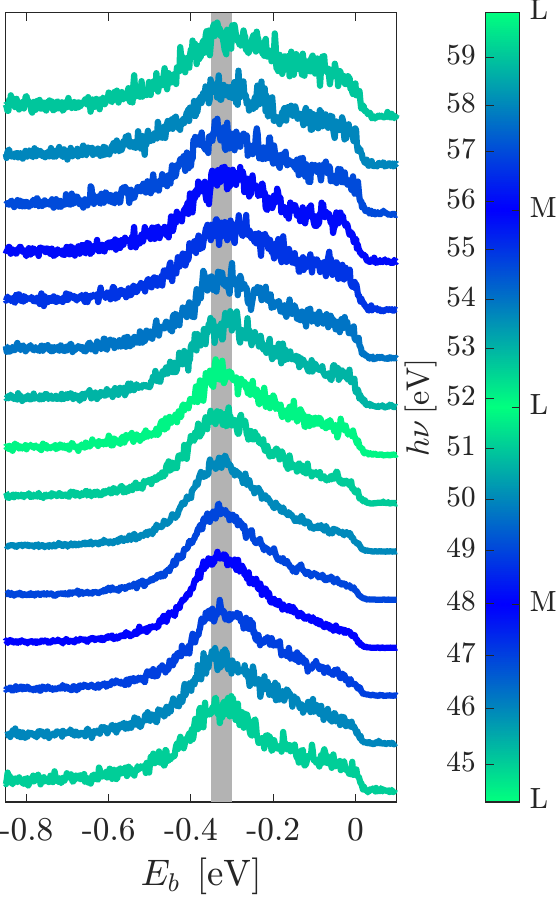}
\caption{EDC at the M point depicted across photon energies ranging from \(45\eV\) to \(59\eV\). Considering \(V_0=9\eV\), this sequence represents movement in the \(k_z\) direction, as indicated by the color bar. The vertical gray line highlights the valence bands' binding energies.}
\label{fig:kzDep}
\end{figure}

\section{\label{sec:conclusions}CONCLUSIONS}
A detailed study of the misfit material $\left(\rm{PbS}\right)_{1.13} \left(\rm{TaS}_2\right)$ reveals the two-dimensional nature of its electronic properties in the bulk form. The transport properties show highly anisotropic superconducting properties with an increased superconducting transition at 3.14 K in comparison to the bulk 2H-TaS$_2$. The elevated transition temperature and lack of charge density wave reflect its similarity to monolayer TaS$_2$, suggesting that the monochalcogenide layer fully decouples the TaS$_2$ layers.  The two-dimensional nature of electronic properties is further demonstrated by the ARPES data. At the Fermi level, the spectra are dominated by the TaS$_2$ layers, while the interlayer coupling between the TaS$_2$ layer is found to be negligible. The spin-resolved ARPES shows strong spin-valley locking, originating from the noncentrosymmetric nature of the TaS$_2$ monolayers and the high spin-orbit coupling. A significant charge transfer effect of about 0.2 electrons per Ta atom is observed from the monochalcogenide to the dichalcogenide layer. The in-plane upper critical field exceeded the conventional Pauli limit, as expected in an Ising superconductor. 

In summary, our findings demonstrate a robust spin-valley polarization in a bulk material, a phenomenon typically seen only in monolayer or few-layer samples. This positions misfit-layered materials as a promising platform for exploring valleytronics while also offering a unique opportunity to study proximity and correlation effects in quantum materials.
 
\section{\label{sec:metods} EXPERIMENTAL METHODS}
Single crystal samples of \(\left(\mathrm{PbS}\right)_{1.13} \left(\mathrm{TaS}_2\right)\) were synthesized by the chemical vapor transport method, with \(\mathrm{PbCl}_2\) serving as the transport agent. For the synthesis, 150 mg of \(\mathrm{PbCl}_2\) was combined with a pre-sintered sample and sealed in a quartz ampule. The ampule was placed in a three-zone furnace and subjected to a temperature gradient of 100$^{\circ}$ C for 14 days. After this period, shiny, silver-colored single crystals were gathered at the colder end of the ampule.

Characterization of the crystals was conducted using X-ray powder diffraction (XRD) with a Rigaku Smartlab 9 kW high-resolution diffraction system employing \(\mathrm{Cu\; K}_{\alpha}\) radiation (\(\lambda = 1.54056 \text{\AA}\)). Cross-sectional high-resolution transmission electron microscopy (HR-TEM) images were acquired using a Titan Themis G2 60-300 (FEI) microscope.

Magnetic properties were assessed via DC magnetic susceptibility measurements performed with a Quantum Design superconducting quantum interference device (QD - MPMS3). Measurements were taken in both zero-field-cooled (ZFC) and field-cooled (FC) modes. Additionally, electrical properties were evaluated using a Quantum Design physical property measurement system (Dynacool-PPMS), with orientation-dependent transport properties examined by rotating the sample within the measurement apparatus. The heat capacity measurements were performed with the PPMS equipped with a dilution refrigerator in the 0.05 K to 4 K temperature range, using the two-tau relaxation method.

We performed transverse-field (TF) $\mu SR$ experiments using the Dolly spectrometer ($πE1$ beamline) at the Paul Scherrer Institute (Villigen, Switzerland). The Dolly spectrometer is equipped with a standard veto setup, providing a low background µSR signal. TF- $\mu SR$ experiment was done after field-cooling the sample with the applied field perpendicular to the $ab$-plane. The $\mu SR$ time spectra were analyzed using the open software package Musrfit \cite{musrfit}.

ARPES measurements were conducted at three different facilities: the SIS - X09LA beamline at the Swiss Light Source (SLS), the CASSIOPEE beamline at the SOLEIL synchrotron, and the spin-ARPES station at the U125-PGM beamline of
the synchrotron source BESSY-II. All ARPES experiments were carried out under ultra-high vacuum (UHV) conditions, maintaining a pressure below \(5\times 10^{-11} \mathrm{torr}\), at an approximate temperature of \(40\K\). The spin-ARPES beamline featured two Mott-type spin detectors, facilitating spin-resolved measurements and allowing for a detailed examination of the electronic band structure and spin characteristics.

\section{Acknowledgments}
We acknowledge useful discussions with Jonathan Ruhman. We thank Hennadii Yerzhakov for providing the parameters of the tight-binding model.
The $\mu$SR measurements were performed at the Swiss Muon Source (S$\mu$S) at the Paul Scherrer Institute in Villigen, Switzerland. 
We acknowledge the Paul Scherrer Institute, Villigen, Switzerland for provision of synchrotron radiation beamtime at beamline SIS of the SLS.  We thank the Helmholtz Zentrum Berlin for the allocation of synchrotron radiation beamtime. We acknowledge SOLEIL for provision of synchrotron radiation facilities. The work at the Technion was supported by Israeli Science Foundation grant number ISF-1263/21.

\bibliographystyle{apsrev4-2}
\bibliography{MainBib}

\begin{thebibliography}{63}%
\makeatletter
\providecommand \@ifxundefined [1]{%
 \@ifx{#1\undefined}
}%
\providecommand \@ifnum [1]{%
 \ifnum #1\expandafter \@firstoftwo
 \else \expandafter \@secondoftwo
 \fi
}%
\providecommand \@ifx [1]{%
 \ifx #1\expandafter \@firstoftwo
 \else \expandafter \@secondoftwo
 \fi
}%
\providecommand \natexlab [1]{#1}%
\providecommand \enquote  [1]{``#1''}%
\providecommand \bibnamefont  [1]{#1}%
\providecommand \bibfnamefont [1]{#1}%
\providecommand \citenamefont [1]{#1}%
\providecommand \href@noop [0]{\@secondoftwo}%
\providecommand \href [0]{\begingroup \@sanitize@url \@href}%
\providecommand \@href[1]{\@@startlink{#1}\@@href}%
\providecommand \@@href[1]{\endgroup#1\@@endlink}%
\providecommand \@sanitize@url [0]{\catcode `\\12\catcode `\$12\catcode
  `\&12\catcode `\#12\catcode `\^12\catcode `\_12\catcode `\%12\relax}%
\providecommand \@@startlink[1]{}%
\providecommand \@@endlink[0]{}%
\providecommand \url  [0]{\begingroup\@sanitize@url \@url }%
\providecommand \@url [1]{\endgroup\@href {#1}{\urlprefix }}%
\providecommand \urlprefix  [0]{URL }%
\providecommand \Eprint [0]{\href }%
\providecommand \doibase [0]{https://doi.org/}%
\providecommand \selectlanguage [0]{\@gobble}%
\providecommand \bibinfo  [0]{\@secondoftwo}%
\providecommand \bibfield  [0]{\@secondoftwo}%
\providecommand \translation [1]{[#1]}%
\providecommand \BibitemOpen [0]{}%
\providecommand \bibitemStop [0]{}%
\providecommand \bibitemNoStop [0]{.\EOS\space}%
\providecommand \EOS [0]{\spacefactor3000\relax}%
\providecommand \BibitemShut  [1]{\csname bibitem#1\endcsname}%
\let\auto@bib@innerbib\@empty
\bibitem [{\citenamefont {Manzeli}\ \emph {et~al.}(2017)\citenamefont
  {Manzeli}, \citenamefont {Ovchinnikov}, \citenamefont {Pasquier},
  \citenamefont {Yazyev},\ and\ \citenamefont {Kis}}]{TMDC}%
  \BibitemOpen
  \bibfield  {author} {\bibinfo {author} {\bibfnamefont {S.}~\bibnamefont
  {Manzeli}}, \bibinfo {author} {\bibfnamefont {D.}~\bibnamefont
  {Ovchinnikov}}, \bibinfo {author} {\bibfnamefont {D.}~\bibnamefont
  {Pasquier}}, \bibinfo {author} {\bibfnamefont {O.~V.}\ \bibnamefont
  {Yazyev}},\ and\ \bibinfo {author} {\bibfnamefont {A.}~\bibnamefont {Kis}},\
  }\href {https://doi.org/https://doi.org/10.1038/natrevmats.2017.33}
  {\bibfield  {journal} {\bibinfo  {journal} {Nature Reviews Materials}\
  }\textbf {\bibinfo {volume} {2}},\ \bibinfo {pages} {1} (\bibinfo {year}
  {2017})}\BibitemShut {NoStop}%
\bibitem [{\citenamefont {Chowdhury}\ \emph {et~al.}(2020)\citenamefont
  {Chowdhury}, \citenamefont {Sadler},\ and\ \citenamefont {Kempa}}]{TMDC2}%
  \BibitemOpen
  \bibfield  {author} {\bibinfo {author} {\bibfnamefont {T.}~\bibnamefont
  {Chowdhury}}, \bibinfo {author} {\bibfnamefont {E.~C.}\ \bibnamefont
  {Sadler}},\ and\ \bibinfo {author} {\bibfnamefont {T.~J.}\ \bibnamefont
  {Kempa}},\ }\href
  {https://doi.org/https://doi.org/10.1021/acs.chemrev.0c00505} {\bibfield
  {journal} {\bibinfo  {journal} {Chemical Reviews}\ }\textbf {\bibinfo
  {volume} {120}},\ \bibinfo {pages} {12563} (\bibinfo {year}
  {2020})}\BibitemShut {NoStop}%
\bibitem [{\citenamefont {Liu}\ \emph {et~al.}(2019)\citenamefont {Liu},
  \citenamefont {Gao}, \citenamefont {He}, \citenamefont {Yu},\ and\
  \citenamefont {Liu}}]{valley}%
  \BibitemOpen
  \bibfield  {author} {\bibinfo {author} {\bibfnamefont {Y.}~\bibnamefont
  {Liu}}, \bibinfo {author} {\bibfnamefont {S.}~\bibnamefont {Gao},
  \bibfnamefont {Yuanji~andZhang}}, \bibinfo {author} {\bibfnamefont
  {J.}~\bibnamefont {He}}, \bibinfo {author} {\bibfnamefont {J.}~\bibnamefont
  {Yu}},\ and\ \bibinfo {author} {\bibfnamefont {Z.}~\bibnamefont {Liu}},\
  }\href {https://doi.org/https://doi.org/10.1007/s12274-019-2497-2} {\bibfield
   {journal} {\bibinfo  {journal} {Nano Research}\ }\textbf {\bibinfo {volume}
  {12}},\ \bibinfo {pages} {2695} (\bibinfo {year} {2019})}\BibitemShut
  {NoStop}%
\bibitem [{\citenamefont {Wang}\ \emph {et~al.}(2015)\citenamefont {Wang},
  \citenamefont {Yuan}, \citenamefont {Hong}, \citenamefont {Li},\ and\
  \citenamefont {Cui}}]{tuning}%
  \BibitemOpen
  \bibfield  {author} {\bibinfo {author} {\bibfnamefont {H.}~\bibnamefont
  {Wang}}, \bibinfo {author} {\bibfnamefont {H.}~\bibnamefont {Yuan}}, \bibinfo
  {author} {\bibfnamefont {S.~S.}\ \bibnamefont {Hong}}, \bibinfo {author}
  {\bibfnamefont {Y.}~\bibnamefont {Li}},\ and\ \bibinfo {author}
  {\bibfnamefont {Y.}~\bibnamefont {Cui}},\ }\href
  {https://doi.org/https://doi.org/10.1039/C4CS00287C} {\bibfield  {journal}
  {\bibinfo  {journal} {Chemical Society Reviews}\ }\textbf {\bibinfo {volume}
  {44}},\ \bibinfo {pages} {2664} (\bibinfo {year} {2015})}\BibitemShut
  {NoStop}%
\bibitem [{\citenamefont {Jung}\ \emph {et~al.}(2016)\citenamefont {Jung},
  \citenamefont {Zhou},\ and\ \citenamefont {Cha}}]{intercalation}%
  \BibitemOpen
  \bibfield  {author} {\bibinfo {author} {\bibfnamefont {Y.}~\bibnamefont
  {Jung}}, \bibinfo {author} {\bibfnamefont {Y.}~\bibnamefont {Zhou}},\ and\
  \bibinfo {author} {\bibfnamefont {J.~J.}\ \bibnamefont {Cha}},\ }\href
  {https://doi.org/DOI https://doi.org/10.1039/C5QI00242G} {\bibfield
  {journal} {\bibinfo  {journal} {Inorganic Chemistry Frontiers}\ }\textbf
  {\bibinfo {volume} {3}},\ \bibinfo {pages} {452} (\bibinfo {year}
  {2016})}\BibitemShut {NoStop}%
\bibitem [{\citenamefont {Tedstone}\ \emph {et~al.}(2016)\citenamefont
  {Tedstone}, \citenamefont {Lewis},\ and\ \citenamefont {O’Brien}}]{doping}%
  \BibitemOpen
  \bibfield  {author} {\bibinfo {author} {\bibfnamefont {A.~A.}\ \bibnamefont
  {Tedstone}}, \bibinfo {author} {\bibfnamefont {D.~J.}\ \bibnamefont
  {Lewis}},\ and\ \bibinfo {author} {\bibfnamefont {P.}~\bibnamefont
  {O’Brien}},\ }\href
  {https://doi.org/https://doi.org/10.1021/acs.chemmater.6b00430} {\bibfield
  {journal} {\bibinfo  {journal} {Chemistry of Materials}\ }\textbf {\bibinfo
  {volume} {28}},\ \bibinfo {pages} {1965} (\bibinfo {year}
  {2016})}\BibitemShut {NoStop}%
\bibitem [{\citenamefont {Shi}\ \emph {et~al.}(2015)\citenamefont {Shi},
  \citenamefont {Ye}, \citenamefont {Zhang}, \citenamefont {Suzuki},
  \citenamefont {Yoshida}, \citenamefont {Miyazaki}, \citenamefont {Inoue},
  \citenamefont {Saito},\ and\ \citenamefont {Iwasa}}]{gating}%
  \BibitemOpen
  \bibfield  {author} {\bibinfo {author} {\bibfnamefont {W.}~\bibnamefont
  {Shi}}, \bibinfo {author} {\bibfnamefont {J.}~\bibnamefont {Ye}}, \bibinfo
  {author} {\bibfnamefont {Y.}~\bibnamefont {Zhang}}, \bibinfo {author}
  {\bibfnamefont {R.}~\bibnamefont {Suzuki}}, \bibinfo {author} {\bibfnamefont
  {M.}~\bibnamefont {Yoshida}}, \bibinfo {author} {\bibfnamefont
  {J.}~\bibnamefont {Miyazaki}}, \bibinfo {author} {\bibfnamefont
  {N.}~\bibnamefont {Inoue}}, \bibinfo {author} {\bibfnamefont
  {Y.}~\bibnamefont {Saito}},\ and\ \bibinfo {author} {\bibfnamefont
  {Y.}~\bibnamefont {Iwasa}},\ }\href
  {https://doi.org/https://doi.org/10.1038/srep12534} {\bibfield  {journal}
  {\bibinfo  {journal} {Scientific reports}\ }\textbf {\bibinfo {volume} {5}},\
  \bibinfo {pages} {12534} (\bibinfo {year} {2015})}\BibitemShut {NoStop}%
\bibitem [{\citenamefont {Novoselov}\ \emph {et~al.}(2016)\citenamefont
  {Novoselov}, \citenamefont {Mishchenko}, \citenamefont {Carvalho},\ and\
  \citenamefont {Neto}}]{hetero}%
  \BibitemOpen
  \bibfield  {author} {\bibinfo {author} {\bibfnamefont {K.~S.}\ \bibnamefont
  {Novoselov}}, \bibinfo {author} {\bibfnamefont {A.}~\bibnamefont
  {Mishchenko}}, \bibinfo {author} {\bibfnamefont {A.}~\bibnamefont
  {Carvalho}},\ and\ \bibinfo {author} {\bibfnamefont {A.~H.~C.}\ \bibnamefont
  {Neto}},\ }\href {https://doi.org/DOI: 10.1126/science.aac9439} {\bibfield
  {journal} {\bibinfo  {journal} {Science}\ }\textbf {\bibinfo {volume}
  {353}},\ \bibinfo {pages} {aac9439} (\bibinfo {year} {2016})}\BibitemShut
  {NoStop}%
\bibitem [{\citenamefont {Liu}\ \emph {et~al.}(2016)\citenamefont {Liu},
  \citenamefont {Weiss}, \citenamefont {Duan}, \citenamefont {Cheng},
  \citenamefont {Huang},\ and\ \citenamefont {Duan}}]{hetero2}%
  \BibitemOpen
  \bibfield  {author} {\bibinfo {author} {\bibfnamefont {Y.}~\bibnamefont
  {Liu}}, \bibinfo {author} {\bibfnamefont {N.~O.}\ \bibnamefont {Weiss}},
  \bibinfo {author} {\bibfnamefont {X.}~\bibnamefont {Duan}}, \bibinfo {author}
  {\bibfnamefont {H.-C.}\ \bibnamefont {Cheng}}, \bibinfo {author}
  {\bibfnamefont {Y.}~\bibnamefont {Huang}},\ and\ \bibinfo {author}
  {\bibfnamefont {X.}~\bibnamefont {Duan}},\ }\href
  {https://doi.org/https://doi.org/10.1038/natrevmats.2016.42} {\bibfield
  {journal} {\bibinfo  {journal} {Nature Reviews Materials}\ }\textbf {\bibinfo
  {volume} {1}},\ \bibinfo {pages} {1} (\bibinfo {year} {2016})}\BibitemShut
  {NoStop}%
\bibitem [{\citenamefont {Trang}\ \emph {et~al.}(2020)\citenamefont {Trang},
  \citenamefont {Shimamura}, \citenamefont {Nakayama}, \citenamefont {Souma},
  \citenamefont {Sugawara}, \citenamefont {Watanabe}, \citenamefont {Yamauchi},
  \citenamefont {Oguchi}, \citenamefont {Segawa}, \citenamefont {Takahashi}
  \emph {et~al.}}]{tsc}%
  \BibitemOpen
  \bibfield  {author} {\bibinfo {author} {\bibfnamefont {C.}~\bibnamefont
  {Trang}}, \bibinfo {author} {\bibfnamefont {N.}~\bibnamefont {Shimamura}},
  \bibinfo {author} {\bibfnamefont {K.}~\bibnamefont {Nakayama}}, \bibinfo
  {author} {\bibfnamefont {S.}~\bibnamefont {Souma}}, \bibinfo {author}
  {\bibfnamefont {K.}~\bibnamefont {Sugawara}}, \bibinfo {author}
  {\bibfnamefont {I.}~\bibnamefont {Watanabe}}, \bibinfo {author}
  {\bibfnamefont {K.}~\bibnamefont {Yamauchi}}, \bibinfo {author}
  {\bibfnamefont {T.}~\bibnamefont {Oguchi}}, \bibinfo {author} {\bibfnamefont
  {K.}~\bibnamefont {Segawa}}, \bibinfo {author} {\bibfnamefont
  {T.}~\bibnamefont {Takahashi}}, \emph {et~al.},\ }\href
  {https://doi.org/https://doi.org/10.1038/s41467-019-13946-0} {\bibfield
  {journal} {\bibinfo  {journal} {Nature communications}\ }\textbf {\bibinfo
  {volume} {11}},\ \bibinfo {pages} {159} (\bibinfo {year} {2020})}\BibitemShut
  {NoStop}%
\bibitem [{\citenamefont {Kezilebieke}\ \emph {et~al.}(2020)\citenamefont
  {Kezilebieke}, \citenamefont {Huda}, \citenamefont {Va{\v{n}}o},
  \citenamefont {Aapro}, \citenamefont {Ganguli}, \citenamefont {Silveira},
  \citenamefont {G{\l}odzik}, \citenamefont {Foster}, \citenamefont {Ojanen},\
  and\ \citenamefont {Liljeroth}}]{tsc2}%
  \BibitemOpen
  \bibfield  {author} {\bibinfo {author} {\bibfnamefont {S.}~\bibnamefont
  {Kezilebieke}}, \bibinfo {author} {\bibfnamefont {M.~N.}\ \bibnamefont
  {Huda}}, \bibinfo {author} {\bibfnamefont {V.}~\bibnamefont {Va{\v{n}}o}},
  \bibinfo {author} {\bibfnamefont {M.}~\bibnamefont {Aapro}}, \bibinfo
  {author} {\bibfnamefont {S.~C.}\ \bibnamefont {Ganguli}}, \bibinfo {author}
  {\bibfnamefont {O.~J.}\ \bibnamefont {Silveira}}, \bibinfo {author}
  {\bibfnamefont {S.}~\bibnamefont {G{\l}odzik}}, \bibinfo {author}
  {\bibfnamefont {A.~S.}\ \bibnamefont {Foster}}, \bibinfo {author}
  {\bibfnamefont {T.}~\bibnamefont {Ojanen}},\ and\ \bibinfo {author}
  {\bibfnamefont {P.}~\bibnamefont {Liljeroth}},\ }\href
  {https://doi.org/https://doi.org/10.1038/s41586-020-2989-y} {\bibfield
  {journal} {\bibinfo  {journal} {Nature}\ }\textbf {\bibinfo {volume} {588}},\
  \bibinfo {pages} {424} (\bibinfo {year} {2020})}\BibitemShut {NoStop}%
\bibitem [{\citenamefont {Lian}\ \emph {et~al.}(2018)\citenamefont {Lian},
  \citenamefont {Sun}, \citenamefont {Vaezi}, \citenamefont {Qi},\ and\
  \citenamefont {Zhang}}]{majorana}%
  \BibitemOpen
  \bibfield  {author} {\bibinfo {author} {\bibfnamefont {B.}~\bibnamefont
  {Lian}}, \bibinfo {author} {\bibfnamefont {X.-Q.}\ \bibnamefont {Sun}},
  \bibinfo {author} {\bibfnamefont {A.}~\bibnamefont {Vaezi}}, \bibinfo
  {author} {\bibfnamefont {X.-L.}\ \bibnamefont {Qi}},\ and\ \bibinfo {author}
  {\bibfnamefont {S.-C.}\ \bibnamefont {Zhang}},\ }\href
  {https://doi.org/https://doi.org/10.1073/pnas.1810003115} {\bibfield
  {journal} {\bibinfo  {journal} {Proceedings of the National Academy of
  Sciences}\ }\textbf {\bibinfo {volume} {115}},\ \bibinfo {pages} {10938}
  (\bibinfo {year} {2018})}\BibitemShut {NoStop}%
\bibitem [{\citenamefont {Zhang}\ \emph {et~al.}(2021)\citenamefont {Zhang},
  \citenamefont {Liu}, \citenamefont {Kirchner}, \citenamefont {Wang},
  \citenamefont {Ren},\ and\ \citenamefont {Lei}}]{issues}%
  \BibitemOpen
  \bibfield  {author} {\bibinfo {author} {\bibfnamefont {S.}~\bibnamefont
  {Zhang}}, \bibinfo {author} {\bibfnamefont {J.}~\bibnamefont {Liu}}, \bibinfo
  {author} {\bibfnamefont {M.~M.}\ \bibnamefont {Kirchner}}, \bibinfo {author}
  {\bibfnamefont {H.}~\bibnamefont {Wang}}, \bibinfo {author} {\bibfnamefont
  {Y.}~\bibnamefont {Ren}},\ and\ \bibinfo {author} {\bibfnamefont
  {W.}~\bibnamefont {Lei}},\ }\href {https://doi.org/10.1088/1361-6463/ac16a4}
  {\bibfield  {journal} {\bibinfo  {journal} {Journal of Physics D: Applied
  Physics}\ }\textbf {\bibinfo {volume} {54}},\ \bibinfo {pages} {433001}
  (\bibinfo {year} {2021})}\BibitemShut {NoStop}%
\bibitem [{\citenamefont {Zhang}\ \emph {et~al.}(2015)\citenamefont {Zhang},
  \citenamefont {Lu}, \citenamefont {Wang}, \citenamefont {Cheng},
  \citenamefont {Zhang},\ and\ \citenamefont {Deng}}]{issues2}%
  \BibitemOpen
  \bibfield  {author} {\bibinfo {author} {\bibfnamefont {H.}~\bibnamefont
  {Zhang}}, \bibinfo {author} {\bibfnamefont {Y.}~\bibnamefont {Lu}}, \bibinfo
  {author} {\bibfnamefont {C.}~\bibnamefont {Wang}}, \bibinfo {author}
  {\bibfnamefont {J.}~\bibnamefont {Cheng}}, \bibinfo {author} {\bibfnamefont
  {Y.}~\bibnamefont {Zhang}},\ and\ \bibinfo {author} {\bibfnamefont
  {Y.}~\bibnamefont {Deng}},\ }\href
  {https://doi.org/10.1021/acs.nanolett.5b00648} {\bibfield  {journal}
  {\bibinfo  {journal} {Nano Letters}\ }\textbf {\bibinfo {volume} {15}},\
  \bibinfo {pages} {3060} (\bibinfo {year} {2015})}\BibitemShut {NoStop}%
\bibitem [{\citenamefont {Ng}\ and\ \citenamefont {McQueen}(2022)}]{misfit1}%
  \BibitemOpen
  \bibfield  {author} {\bibinfo {author} {\bibfnamefont {N.}~\bibnamefont
  {Ng}}\ and\ \bibinfo {author} {\bibfnamefont {T.~M.}\ \bibnamefont
  {McQueen}},\ }\bibfield  {journal} {\bibinfo  {journal} {APL Materials}\
  }\textbf {\bibinfo {volume} {10}},\ \href
  {https://doi.org/https://doi.org/10.1063/5.0101429}
  {https://doi.org/10.1063/5.0101429} (\bibinfo {year} {2022})\BibitemShut
  {NoStop}%
\bibitem [{\citenamefont {Leriche}\ \emph
  {et~al.}(2021{\natexlab{a}})\citenamefont {Leriche}, \citenamefont
  {Palacio-Morales}, \citenamefont {Campetella}, \citenamefont {Tresca},
  \citenamefont {Sasaki}, \citenamefont {Brun}, \citenamefont {Debontridder},
  \citenamefont {David}, \citenamefont {Arfaoui}, \citenamefont
  {{\v{S}}ofranko} \emph {et~al.}}]{misfit2}%
  \BibitemOpen
  \bibfield  {author} {\bibinfo {author} {\bibfnamefont {R.~T.}\ \bibnamefont
  {Leriche}}, \bibinfo {author} {\bibfnamefont {A.}~\bibnamefont
  {Palacio-Morales}}, \bibinfo {author} {\bibfnamefont {M.}~\bibnamefont
  {Campetella}}, \bibinfo {author} {\bibfnamefont {C.}~\bibnamefont {Tresca}},
  \bibinfo {author} {\bibfnamefont {S.}~\bibnamefont {Sasaki}}, \bibinfo
  {author} {\bibfnamefont {C.}~\bibnamefont {Brun}}, \bibinfo {author}
  {\bibfnamefont {F.}~\bibnamefont {Debontridder}}, \bibinfo {author}
  {\bibfnamefont {P.}~\bibnamefont {David}}, \bibinfo {author} {\bibfnamefont
  {I.}~\bibnamefont {Arfaoui}}, \bibinfo {author} {\bibfnamefont
  {O.}~\bibnamefont {{\v{S}}ofranko}}, \emph {et~al.},\ }\href
  {https://doi.org/https://doi.org/10.1002/adfm.202007706} {\bibfield
  {journal} {\bibinfo  {journal} {Advanced Functional Materials}\ }\textbf
  {\bibinfo {volume} {31}},\ \bibinfo {pages} {2007706} (\bibinfo {year}
  {2021}{\natexlab{a}})}\BibitemShut {NoStop}%
\bibitem [{\citenamefont {Dolotko}\ \emph {et~al.}(2020)\citenamefont
  {Dolotko}, \citenamefont {Hlova}, \citenamefont {Pathak}, \citenamefont
  {Mudryk}, \citenamefont {Pecharsky}, \citenamefont {Singh}, \citenamefont
  {Johnson}, \citenamefont {Boote}, \citenamefont {Li}, \citenamefont {Smith}
  \emph {et~al.}}]{misfit3}%
  \BibitemOpen
  \bibfield  {author} {\bibinfo {author} {\bibfnamefont {O.}~\bibnamefont
  {Dolotko}}, \bibinfo {author} {\bibfnamefont {I.~Z.}\ \bibnamefont {Hlova}},
  \bibinfo {author} {\bibfnamefont {A.~K.}\ \bibnamefont {Pathak}}, \bibinfo
  {author} {\bibfnamefont {Y.}~\bibnamefont {Mudryk}}, \bibinfo {author}
  {\bibfnamefont {V.~K.}\ \bibnamefont {Pecharsky}}, \bibinfo {author}
  {\bibfnamefont {P.}~\bibnamefont {Singh}}, \bibinfo {author} {\bibfnamefont
  {D.~D.}\ \bibnamefont {Johnson}}, \bibinfo {author} {\bibfnamefont {B.~W.}\
  \bibnamefont {Boote}}, \bibinfo {author} {\bibfnamefont {J.}~\bibnamefont
  {Li}}, \bibinfo {author} {\bibfnamefont {E.~A.}\ \bibnamefont {Smith}}, \emph
  {et~al.},\ }\href
  {https://doi.org/https://doi.org/10.1038/s41467-020-16672-0} {\bibfield
  {journal} {\bibinfo  {journal} {Nature Communications}\ }\textbf {\bibinfo
  {volume} {11}},\ \bibinfo {pages} {3005} (\bibinfo {year}
  {2020})}\BibitemShut {NoStop}%
\bibitem [{\citenamefont {Brandt}\ \emph {et~al.}(2003)\citenamefont {Brandt},
  \citenamefont {Kipp}, \citenamefont {Skibowski}, \citenamefont {Krasovskii},
  \citenamefont {Schattke}, \citenamefont {Spiecker}, \citenamefont {Dieker},\
  and\ \citenamefont {J{\"a}ger}}]{charge1}%
  \BibitemOpen
  \bibfield  {author} {\bibinfo {author} {\bibfnamefont {J.}~\bibnamefont
  {Brandt}}, \bibinfo {author} {\bibfnamefont {L.}~\bibnamefont {Kipp}},
  \bibinfo {author} {\bibfnamefont {M.}~\bibnamefont {Skibowski}}, \bibinfo
  {author} {\bibfnamefont {E.}~\bibnamefont {Krasovskii}}, \bibinfo {author}
  {\bibfnamefont {W.}~\bibnamefont {Schattke}}, \bibinfo {author}
  {\bibfnamefont {E.}~\bibnamefont {Spiecker}}, \bibinfo {author}
  {\bibfnamefont {C.}~\bibnamefont {Dieker}},\ and\ \bibinfo {author}
  {\bibfnamefont {W.}~\bibnamefont {J{\"a}ger}},\ }\href
  {https://doi.org/https://doi.org/10.1016/S0039-6028(03)00165-1} {\bibfield
  {journal} {\bibinfo  {journal} {Surface science}\ }\textbf {\bibinfo {volume}
  {532}},\ \bibinfo {pages} {705} (\bibinfo {year} {2003})}\BibitemShut
  {NoStop}%
\bibitem [{\citenamefont {Samuely}\ \emph {et~al.}(2023)\citenamefont
  {Samuely}, \citenamefont {Wickramaratne}, \citenamefont {Gmitra},
  \citenamefont {Jaouen}, \citenamefont {\ifmmode~\check{S}\else
  \v{S}\fi{}ofranko}, \citenamefont {Volavka}, \citenamefont {Kuzmiak},
  \citenamefont {Hani\ifmmode~\check{s}\else \v{s}\fi{}}, \citenamefont
  {Szab\'o}, \citenamefont {Monney}, \citenamefont {Kremer}, \citenamefont
  {Le~F\`evre}, \citenamefont {Bertran}, \citenamefont {Cren}, \citenamefont
  {Sasaki}, \citenamefont {Cario}, \citenamefont {Calandra}, \citenamefont
  {Mazin},\ and\ \citenamefont {Samuely}}]{charge2}%
  \BibitemOpen
  \bibfield  {author} {\bibinfo {author} {\bibfnamefont {T.}~\bibnamefont
  {Samuely}}, \bibinfo {author} {\bibfnamefont {D.}~\bibnamefont
  {Wickramaratne}}, \bibinfo {author} {\bibfnamefont {M.}~\bibnamefont
  {Gmitra}}, \bibinfo {author} {\bibfnamefont {T.}~\bibnamefont {Jaouen}},
  \bibinfo {author} {\bibfnamefont {O.}~\bibnamefont {\ifmmode~\check{S}\else
  \v{S}\fi{}ofranko}}, \bibinfo {author} {\bibfnamefont {D.}~\bibnamefont
  {Volavka}}, \bibinfo {author} {\bibfnamefont {M.}~\bibnamefont {Kuzmiak}},
  \bibinfo {author} {\bibfnamefont {J.}~\bibnamefont
  {Hani\ifmmode~\check{s}\else \v{s}\fi{}}}, \bibinfo {author} {\bibfnamefont
  {P.}~\bibnamefont {Szab\'o}}, \bibinfo {author} {\bibfnamefont
  {C.}~\bibnamefont {Monney}}, \bibinfo {author} {\bibfnamefont
  {G.}~\bibnamefont {Kremer}}, \bibinfo {author} {\bibfnamefont
  {P.}~\bibnamefont {Le~F\`evre}}, \bibinfo {author} {\bibfnamefont {F.~m.~c.}\
  \bibnamefont {Bertran}}, \bibinfo {author} {\bibfnamefont {T.}~\bibnamefont
  {Cren}}, \bibinfo {author} {\bibfnamefont {S.}~\bibnamefont {Sasaki}},
  \bibinfo {author} {\bibfnamefont {L.}~\bibnamefont {Cario}}, \bibinfo
  {author} {\bibfnamefont {M.}~\bibnamefont {Calandra}}, \bibinfo {author}
  {\bibfnamefont {I.~I.}\ \bibnamefont {Mazin}},\ and\ \bibinfo {author}
  {\bibfnamefont {P.}~\bibnamefont {Samuely}},\ }\href
  {https://doi.org/10.1103/PhysRevB.108.L220501} {\bibfield  {journal}
  {\bibinfo  {journal} {Phys. Rev. B}\ }\textbf {\bibinfo {volume} {108}},\
  \bibinfo {pages} {L220501} (\bibinfo {year} {2023})}\BibitemShut {NoStop}%
\bibitem [{\citenamefont {Yang}\ \emph
  {et~al.}(2018{\natexlab{a}})\citenamefont {Yang}, \citenamefont {Wang},
  \citenamefont {Li}, \citenamefont {Bai}, \citenamefont {Ma}, \citenamefont
  {Sun}, \citenamefont {Tao}, \citenamefont {Dong},\ and\ \citenamefont
  {Xu}}]{misfit4}%
  \BibitemOpen
  \bibfield  {author} {\bibinfo {author} {\bibfnamefont {X.}~\bibnamefont
  {Yang}}, \bibinfo {author} {\bibfnamefont {M.}~\bibnamefont {Wang}}, \bibinfo
  {author} {\bibfnamefont {Y.}~\bibnamefont {Li}}, \bibinfo {author}
  {\bibfnamefont {H.}~\bibnamefont {Bai}}, \bibinfo {author} {\bibfnamefont
  {J.}~\bibnamefont {Ma}}, \bibinfo {author} {\bibfnamefont {X.}~\bibnamefont
  {Sun}}, \bibinfo {author} {\bibfnamefont {Q.}~\bibnamefont {Tao}}, \bibinfo
  {author} {\bibfnamefont {C.}~\bibnamefont {Dong}},\ and\ \bibinfo {author}
  {\bibfnamefont {Z.-A.}\ \bibnamefont {Xu}},\ }\href
  {https://doi.org/10.1088/1361-6668/aae7b6} {\bibfield  {journal} {\bibinfo
  {journal} {Superconductor Science and Technology}\ }\textbf {\bibinfo
  {volume} {31}},\ \bibinfo {pages} {125010} (\bibinfo {year}
  {2018}{\natexlab{a}})}\BibitemShut {NoStop}%
\bibitem [{\citenamefont {Sankar}\ \emph {et~al.}(2018)\citenamefont {Sankar},
  \citenamefont {Peramaiyan}, \citenamefont {Panneer~Muthuselvam},
  \citenamefont {Wen}, \citenamefont {Xu},\ and\ \citenamefont
  {Chou}}]{misfit5}%
  \BibitemOpen
  \bibfield  {author} {\bibinfo {author} {\bibfnamefont {R.}~\bibnamefont
  {Sankar}}, \bibinfo {author} {\bibfnamefont {G.}~\bibnamefont {Peramaiyan}},
  \bibinfo {author} {\bibfnamefont {I.}~\bibnamefont {Panneer~Muthuselvam}},
  \bibinfo {author} {\bibfnamefont {C.-Y.}\ \bibnamefont {Wen}}, \bibinfo
  {author} {\bibfnamefont {X.}~\bibnamefont {Xu}},\ and\ \bibinfo {author}
  {\bibfnamefont {F.}~\bibnamefont {Chou}},\ }\href
  {https://doi.org/https://doi.org/10.1021/acs.chemmater.7b04998} {\bibfield
  {journal} {\bibinfo  {journal} {Chemistry of Materials}\ }\textbf {\bibinfo
  {volume} {30}},\ \bibinfo {pages} {1373} (\bibinfo {year}
  {2018})}\BibitemShut {NoStop}%
\bibitem [{\citenamefont {Yang}\ \emph {et~al.}(2019)\citenamefont {Yang},
  \citenamefont {Ma}, \citenamefont {Lv}, \citenamefont {Hu}, \citenamefont
  {Sun}, \citenamefont {Li}, \citenamefont {Qiao}, \citenamefont {Wu},
  \citenamefont {Tao}, \citenamefont {Cao} \emph {et~al.}}]{misfit6}%
  \BibitemOpen
  \bibfield  {author} {\bibinfo {author} {\bibfnamefont {X.}~\bibnamefont
  {Yang}}, \bibinfo {author} {\bibfnamefont {J.}~\bibnamefont {Ma}}, \bibinfo
  {author} {\bibfnamefont {B.}~\bibnamefont {Lv}}, \bibinfo {author}
  {\bibfnamefont {H.}~\bibnamefont {Hu}}, \bibinfo {author} {\bibfnamefont
  {T.}~\bibnamefont {Sun}}, \bibinfo {author} {\bibfnamefont {M.}~\bibnamefont
  {Li}}, \bibinfo {author} {\bibfnamefont {L.}~\bibnamefont {Qiao}}, \bibinfo
  {author} {\bibfnamefont {S.}~\bibnamefont {Wu}}, \bibinfo {author}
  {\bibfnamefont {Q.}~\bibnamefont {Tao}}, \bibinfo {author} {\bibfnamefont
  {G.-H.}\ \bibnamefont {Cao}}, \emph {et~al.},\ }\href
  {https://doi.org/10.1209/0295-5075/128/17004} {\bibfield  {journal} {\bibinfo
   {journal} {EPL}\ }\textbf {\bibinfo {volume} {128}},\ \bibinfo {pages}
  {17004} (\bibinfo {year} {2019})}\BibitemShut {NoStop}%
\bibitem [{\citenamefont {Samuely}\ \emph {et~al.}(2021)\citenamefont
  {Samuely}, \citenamefont {Szab\'o}, \citenamefont {Ka\ifmmode \check{c}\else
  \v{c}\fi{}mar\ifmmode~\check{c}\else \v{c}\fi{}\'{\i}k}, \citenamefont
  {Meerschaut}, \citenamefont {Cario}, \citenamefont {Jansen}, \citenamefont
  {Cren}, \citenamefont {Kuzmiak}, \citenamefont {\ifmmode~\check{S}\else
  \v{S}\fi{}ofranko},\ and\ \citenamefont {Samuely}}]{misfit7}%
  \BibitemOpen
  \bibfield  {author} {\bibinfo {author} {\bibfnamefont {P.}~\bibnamefont
  {Samuely}}, \bibinfo {author} {\bibfnamefont {P.}~\bibnamefont {Szab\'o}},
  \bibinfo {author} {\bibfnamefont {J.}~\bibnamefont {Ka\ifmmode \check{c}\else
  \v{c}\fi{}mar\ifmmode~\check{c}\else \v{c}\fi{}\'{\i}k}}, \bibinfo {author}
  {\bibfnamefont {A.}~\bibnamefont {Meerschaut}}, \bibinfo {author}
  {\bibfnamefont {L.}~\bibnamefont {Cario}}, \bibinfo {author} {\bibfnamefont
  {A.~G.~M.}\ \bibnamefont {Jansen}}, \bibinfo {author} {\bibfnamefont
  {T.}~\bibnamefont {Cren}}, \bibinfo {author} {\bibfnamefont {M.}~\bibnamefont
  {Kuzmiak}}, \bibinfo {author} {\bibfnamefont {O.}~\bibnamefont
  {\ifmmode~\check{S}\else \v{S}\fi{}ofranko}},\ and\ \bibinfo {author}
  {\bibfnamefont {T.}~\bibnamefont {Samuely}},\ }\href
  {https://doi.org/10.1103/PhysRevB.104.224507} {\bibfield  {journal} {\bibinfo
   {journal} {Phys. Rev. B}\ }\textbf {\bibinfo {volume} {104}},\ \bibinfo
  {pages} {224507} (\bibinfo {year} {2021})}\BibitemShut {NoStop}%
\bibitem [{\citenamefont {Hummer}\ \emph {et~al.}(2007)\citenamefont {Hummer},
  \citenamefont {Gr\"uneis},\ and\ \citenamefont {Kresse}}]{PbS}%
  \BibitemOpen
  \bibfield  {author} {\bibinfo {author} {\bibfnamefont {K.}~\bibnamefont
  {Hummer}}, \bibinfo {author} {\bibfnamefont {A.}~\bibnamefont {Gr\"uneis}},\
  and\ \bibinfo {author} {\bibfnamefont {G.}~\bibnamefont {Kresse}},\ }\href
  {https://doi.org/10.1103/PhysRevB.75.195211} {\bibfield  {journal} {\bibinfo
  {journal} {Phys. Rev. B}\ }\textbf {\bibinfo {volume} {75}},\ \bibinfo
  {pages} {195211} (\bibinfo {year} {2007})}\BibitemShut {NoStop}%
\bibitem [{\citenamefont {Nagata}\ \emph {et~al.}(1992)\citenamefont {Nagata},
  \citenamefont {Aochi}, \citenamefont {Abe}, \citenamefont {Ebisu},
  \citenamefont {Hagino}, \citenamefont {Seki},\ and\ \citenamefont
  {Tsutsumi}}]{TaS2}%
  \BibitemOpen
  \bibfield  {author} {\bibinfo {author} {\bibfnamefont {S.}~\bibnamefont
  {Nagata}}, \bibinfo {author} {\bibfnamefont {T.}~\bibnamefont {Aochi}},
  \bibinfo {author} {\bibfnamefont {T.}~\bibnamefont {Abe}}, \bibinfo {author}
  {\bibfnamefont {S.}~\bibnamefont {Ebisu}}, \bibinfo {author} {\bibfnamefont
  {T.}~\bibnamefont {Hagino}}, \bibinfo {author} {\bibfnamefont
  {Y.}~\bibnamefont {Seki}},\ and\ \bibinfo {author} {\bibfnamefont
  {K.}~\bibnamefont {Tsutsumi}},\ }\href
  {https://doi.org/10.1016/0022-3697(92)90242-6} {\bibfield  {journal}
  {\bibinfo  {journal} {J. Phys. Chem. Solids}\ }\textbf {\bibinfo {volume}
  {53}},\ \bibinfo {pages} {1259} (\bibinfo {year} {1992})}\BibitemShut
  {NoStop}%
\bibitem [{\citenamefont {Ribak}\ \emph {et~al.}(2020)\citenamefont {Ribak},
  \citenamefont {Skiff}, \citenamefont {Mograbi}, \citenamefont {Rout},
  \citenamefont {Fischer}, \citenamefont {Ruhman}, \citenamefont {Chashka},
  \citenamefont {Dagan},\ and\ \citenamefont {Kanigel}}]{4hb1}%
  \BibitemOpen
  \bibfield  {author} {\bibinfo {author} {\bibfnamefont {A.}~\bibnamefont
  {Ribak}}, \bibinfo {author} {\bibfnamefont {R.~M.}\ \bibnamefont {Skiff}},
  \bibinfo {author} {\bibfnamefont {M.}~\bibnamefont {Mograbi}}, \bibinfo
  {author} {\bibfnamefont {P.}~\bibnamefont {Rout}}, \bibinfo {author}
  {\bibfnamefont {M.}~\bibnamefont {Fischer}}, \bibinfo {author} {\bibfnamefont
  {J.}~\bibnamefont {Ruhman}}, \bibinfo {author} {\bibfnamefont
  {K.}~\bibnamefont {Chashka}}, \bibinfo {author} {\bibfnamefont
  {Y.}~\bibnamefont {Dagan}},\ and\ \bibinfo {author} {\bibfnamefont
  {A.}~\bibnamefont {Kanigel}},\ }\href {https://doi.org/DOI:
  10.1126/sciadv.aax9480} {\bibfield  {journal} {\bibinfo  {journal} {Science
  advances}\ }\textbf {\bibinfo {volume} {6}},\ \bibinfo {pages} {eaax9480}
  (\bibinfo {year} {2020})}\BibitemShut {NoStop}%
\bibitem [{\citenamefont {Silber}\ \emph {et~al.}(2024)\citenamefont {Silber},
  \citenamefont {Mathimalar}, \citenamefont {Mangel}, \citenamefont {Nayak},
  \citenamefont {Green}, \citenamefont {Avraham}, \citenamefont {Beidenkopf},
  \citenamefont {Feldman}, \citenamefont {Kanigel}, \citenamefont {Klein} \emph
  {et~al.}}]{4hb2}%
  \BibitemOpen
  \bibfield  {author} {\bibinfo {author} {\bibfnamefont {I.}~\bibnamefont
  {Silber}}, \bibinfo {author} {\bibfnamefont {S.}~\bibnamefont {Mathimalar}},
  \bibinfo {author} {\bibfnamefont {I.}~\bibnamefont {Mangel}}, \bibinfo
  {author} {\bibfnamefont {A.}~\bibnamefont {Nayak}}, \bibinfo {author}
  {\bibfnamefont {O.}~\bibnamefont {Green}}, \bibinfo {author} {\bibfnamefont
  {N.}~\bibnamefont {Avraham}}, \bibinfo {author} {\bibfnamefont
  {H.}~\bibnamefont {Beidenkopf}}, \bibinfo {author} {\bibfnamefont
  {I.}~\bibnamefont {Feldman}}, \bibinfo {author} {\bibfnamefont
  {A.}~\bibnamefont {Kanigel}}, \bibinfo {author} {\bibfnamefont
  {A.}~\bibnamefont {Klein}}, \emph {et~al.},\ }\href
  {https://doi.org/https://doi.org/10.1038/s41467-024-45169-3} {\bibfield
  {journal} {\bibinfo  {journal} {Nature Communications}\ }\textbf {\bibinfo
  {volume} {15}},\ \bibinfo {pages} {824} (\bibinfo {year} {2024})}\BibitemShut
  {NoStop}%
\bibitem [{\citenamefont {Nayak}\ \emph {et~al.}(2021)\citenamefont {Nayak},
  \citenamefont {Steinbok}, \citenamefont {Roet}, \citenamefont {Koo},
  \citenamefont {Margalit}, \citenamefont {Feldman}, \citenamefont {Almoalem},
  \citenamefont {Kanigel}, \citenamefont {Fiete}, \citenamefont {Yan} \emph
  {et~al.}}]{4hb3}%
  \BibitemOpen
  \bibfield  {author} {\bibinfo {author} {\bibfnamefont {A.~K.}\ \bibnamefont
  {Nayak}}, \bibinfo {author} {\bibfnamefont {A.}~\bibnamefont {Steinbok}},
  \bibinfo {author} {\bibfnamefont {Y.}~\bibnamefont {Roet}}, \bibinfo {author}
  {\bibfnamefont {J.}~\bibnamefont {Koo}}, \bibinfo {author} {\bibfnamefont
  {G.}~\bibnamefont {Margalit}}, \bibinfo {author} {\bibfnamefont
  {I.}~\bibnamefont {Feldman}}, \bibinfo {author} {\bibfnamefont
  {A.}~\bibnamefont {Almoalem}}, \bibinfo {author} {\bibfnamefont
  {A.}~\bibnamefont {Kanigel}}, \bibinfo {author} {\bibfnamefont {G.~A.}\
  \bibnamefont {Fiete}}, \bibinfo {author} {\bibfnamefont {B.}~\bibnamefont
  {Yan}}, \emph {et~al.},\ }\href
  {https://doi.org/https://doi.org/10.1038/s41567-021-01376-z} {\bibfield
  {journal} {\bibinfo  {journal} {Nature physics}\ }\textbf {\bibinfo {volume}
  {17}},\ \bibinfo {pages} {1413} (\bibinfo {year} {2021})}\BibitemShut
  {NoStop}%
\bibitem [{\citenamefont {Almoalem}\ \emph {et~al.}(2024)\citenamefont
  {Almoalem}, \citenamefont {Gofman}, \citenamefont {Nitzav}, \citenamefont
  {Mangel}, \citenamefont {Feldman}, \citenamefont {Koo}, \citenamefont
  {Mazzola}, \citenamefont {Fujii}, \citenamefont {Vobornik}, \citenamefont
  {S{\'{}}~anchez Barriga} \emph {et~al.}}]{4hb4}%
  \BibitemOpen
  \bibfield  {author} {\bibinfo {author} {\bibfnamefont {A.}~\bibnamefont
  {Almoalem}}, \bibinfo {author} {\bibfnamefont {R.}~\bibnamefont {Gofman}},
  \bibinfo {author} {\bibfnamefont {Y.}~\bibnamefont {Nitzav}}, \bibinfo
  {author} {\bibfnamefont {I.}~\bibnamefont {Mangel}}, \bibinfo {author}
  {\bibfnamefont {I.}~\bibnamefont {Feldman}}, \bibinfo {author} {\bibfnamefont
  {J.}~\bibnamefont {Koo}}, \bibinfo {author} {\bibfnamefont {F.}~\bibnamefont
  {Mazzola}}, \bibinfo {author} {\bibfnamefont {J.}~\bibnamefont {Fujii}},
  \bibinfo {author} {\bibfnamefont {I.}~\bibnamefont {Vobornik}}, \bibinfo
  {author} {\bibfnamefont {J.}~\bibnamefont {S{\'{}}~anchez Barriga}}, \emph
  {et~al.},\ }\href
  {https://doi.org/https://doi.org/10.1038/s41535-024-00646-2} {\bibfield
  {journal} {\bibinfo  {journal} {npj Quantum Materials}\ }\textbf {\bibinfo
  {volume} {9}},\ \bibinfo {pages} {36} (\bibinfo {year} {2024})}\BibitemShut
  {NoStop}%
\bibitem [{\citenamefont {Wulff}\ \emph {et~al.}(1990)\citenamefont {Wulff},
  \citenamefont {Meetsma}, \citenamefont {Smaaalen}, \citenamefont {Haange},
  \citenamefont {Boer},\ and\ \citenamefont {Wiegners}}]{PTSxrd}%
  \BibitemOpen
  \bibfield  {author} {\bibinfo {author} {\bibfnamefont {J.}~\bibnamefont
  {Wulff}}, \bibinfo {author} {\bibfnamefont {A.}~\bibnamefont {Meetsma}},
  \bibinfo {author} {\bibfnamefont {S.~V.}\ \bibnamefont {Smaaalen}}, \bibinfo
  {author} {\bibfnamefont {R.~J.}\ \bibnamefont {Haange}}, \bibinfo {author}
  {\bibfnamefont {J.~L.~D.}\ \bibnamefont {Boer}},\ and\ \bibinfo {author}
  {\bibfnamefont {G.~A.}\ \bibnamefont {Wiegners}},\ }\href
  {https://doi.org/https://doi.org/10.1016/0022-4596(90)90190-9} {\bibfield
  {journal} {\bibinfo  {journal} {Journal of solid state chemistry}\ }\textbf
  {\bibinfo {volume} {84}},\ \bibinfo {pages} {118} (\bibinfo {year}
  {1990})}\BibitemShut {NoStop}%
\bibitem [{\citenamefont {Helfand}\ and\ \citenamefont
  {Werthamer}(1966)}]{WHH}%
  \BibitemOpen
  \bibfield  {author} {\bibinfo {author} {\bibfnamefont {E.}~\bibnamefont
  {Helfand}}\ and\ \bibinfo {author} {\bibfnamefont {N.}~\bibnamefont
  {Werthamer}},\ }\href
  {https://doi.org/https://doi.org/10.1103/PhysRev.147.295} {\bibfield
  {journal} {\bibinfo  {journal} {Physical Review}\ }\textbf {\bibinfo {volume}
  {147}},\ \bibinfo {pages} {288} (\bibinfo {year} {1966})}\BibitemShut
  {NoStop}%
\bibitem [{\citenamefont {Navarro-Moratalla}\ \emph {et~al.}(2016)\citenamefont
  {Navarro-Moratalla}, \citenamefont {Island}, \citenamefont {Manas-Valero},
  \citenamefont {Pinilla-Cienfuegos}, \citenamefont {Castellanos-Gomez},
  \citenamefont {Quereda}, \citenamefont {Rubio-Bollinger}, \citenamefont
  {Chirolli}, \citenamefont {Silva-Guill{\'e}n}, \citenamefont {Agra{\"\i}t}
  \emph {et~al.}}]{TaS2mono1}%
  \BibitemOpen
  \bibfield  {author} {\bibinfo {author} {\bibfnamefont {E.}~\bibnamefont
  {Navarro-Moratalla}}, \bibinfo {author} {\bibfnamefont {J.~O.}\ \bibnamefont
  {Island}}, \bibinfo {author} {\bibfnamefont {S.}~\bibnamefont
  {Manas-Valero}}, \bibinfo {author} {\bibfnamefont {E.}~\bibnamefont
  {Pinilla-Cienfuegos}}, \bibinfo {author} {\bibfnamefont {A.}~\bibnamefont
  {Castellanos-Gomez}}, \bibinfo {author} {\bibfnamefont {J.}~\bibnamefont
  {Quereda}}, \bibinfo {author} {\bibfnamefont {G.}~\bibnamefont
  {Rubio-Bollinger}}, \bibinfo {author} {\bibfnamefont {L.}~\bibnamefont
  {Chirolli}}, \bibinfo {author} {\bibfnamefont {J.~A.}\ \bibnamefont
  {Silva-Guill{\'e}n}}, \bibinfo {author} {\bibfnamefont {N.}~\bibnamefont
  {Agra{\"\i}t}}, \emph {et~al.},\ }\href
  {https://doi.org/https://doi.org/10.1038/ncomms11043} {\bibfield  {journal}
  {\bibinfo  {journal} {Nature communications}\ }\textbf {\bibinfo {volume}
  {7}},\ \bibinfo {pages} {11043} (\bibinfo {year} {2016})}\BibitemShut
  {NoStop}%
\bibitem [{\citenamefont {Yang}\ \emph
  {et~al.}(2018{\natexlab{b}})\citenamefont {Yang}, \citenamefont {Fang},
  \citenamefont {Fatemi}, \citenamefont {Ruhman}, \citenamefont
  {Navarro-Moratalla}, \citenamefont {Watanabe}, \citenamefont {Taniguchi},
  \citenamefont {Kaxiras},\ and\ \citenamefont {Jarillo-Herrero}}]{TaS2mono2}%
  \BibitemOpen
  \bibfield  {author} {\bibinfo {author} {\bibfnamefont {Y.}~\bibnamefont
  {Yang}}, \bibinfo {author} {\bibfnamefont {S.}~\bibnamefont {Fang}}, \bibinfo
  {author} {\bibfnamefont {V.}~\bibnamefont {Fatemi}}, \bibinfo {author}
  {\bibfnamefont {J.}~\bibnamefont {Ruhman}}, \bibinfo {author} {\bibfnamefont
  {E.}~\bibnamefont {Navarro-Moratalla}}, \bibinfo {author} {\bibfnamefont
  {K.}~\bibnamefont {Watanabe}}, \bibinfo {author} {\bibfnamefont
  {T.}~\bibnamefont {Taniguchi}}, \bibinfo {author} {\bibfnamefont
  {E.}~\bibnamefont {Kaxiras}},\ and\ \bibinfo {author} {\bibfnamefont
  {P.}~\bibnamefont {Jarillo-Herrero}},\ }\href
  {https://doi.org/https://doi.org/10.1103/PhysRevB.98.035203} {\bibfield
  {journal} {\bibinfo  {journal} {Physical Review B}\ }\textbf {\bibinfo
  {volume} {98}},\ \bibinfo {pages} {035203} (\bibinfo {year}
  {2018}{\natexlab{b}})}\BibitemShut {NoStop}%
\bibitem [{\citenamefont {Xi}\ \emph {et~al.}(2016)\citenamefont {Xi},
  \citenamefont {Wang}, \citenamefont {Zhao}, \citenamefont {Park},
  \citenamefont {Law}, \citenamefont {Berger}, \citenamefont {Forr{\'o}},
  \citenamefont {Shan},\ and\ \citenamefont {Mak}}]{NbSe22}%
  \BibitemOpen
  \bibfield  {author} {\bibinfo {author} {\bibfnamefont {X.}~\bibnamefont
  {Xi}}, \bibinfo {author} {\bibfnamefont {Z.}~\bibnamefont {Wang}}, \bibinfo
  {author} {\bibfnamefont {W.}~\bibnamefont {Zhao}}, \bibinfo {author}
  {\bibfnamefont {J.-H.}\ \bibnamefont {Park}}, \bibinfo {author}
  {\bibfnamefont {K.~T.}\ \bibnamefont {Law}}, \bibinfo {author} {\bibfnamefont
  {H.}~\bibnamefont {Berger}}, \bibinfo {author} {\bibfnamefont
  {L.}~\bibnamefont {Forr{\'o}}}, \bibinfo {author} {\bibfnamefont
  {J.}~\bibnamefont {Shan}},\ and\ \bibinfo {author} {\bibfnamefont {K.~F.}\
  \bibnamefont {Mak}},\ }\href
  {https://doi.org/https://doi.org/10.1038/nphys3538} {\bibfield  {journal}
  {\bibinfo  {journal} {Nature Physics}\ }\textbf {\bibinfo {volume} {12}},\
  \bibinfo {pages} {139} (\bibinfo {year} {2016})}\BibitemShut {NoStop}%
\bibitem [{\citenamefont {Lu}\ \emph {et~al.}(2015)\citenamefont {Lu},
  \citenamefont {Zheliuk}, \citenamefont {Leermakers}, \citenamefont {Yuan},
  \citenamefont {Zeitler}, \citenamefont {Law},\ and\ \citenamefont
  {Ye}}]{mos2}%
  \BibitemOpen
  \bibfield  {author} {\bibinfo {author} {\bibfnamefont {J.}~\bibnamefont
  {Lu}}, \bibinfo {author} {\bibfnamefont {O.}~\bibnamefont {Zheliuk}},
  \bibinfo {author} {\bibfnamefont {I.}~\bibnamefont {Leermakers}}, \bibinfo
  {author} {\bibfnamefont {N.~F.}\ \bibnamefont {Yuan}}, \bibinfo {author}
  {\bibfnamefont {U.}~\bibnamefont {Zeitler}}, \bibinfo {author} {\bibfnamefont
  {K.~T.}\ \bibnamefont {Law}},\ and\ \bibinfo {author} {\bibfnamefont
  {J.}~\bibnamefont {Ye}},\ }\href {https://doi.org/DOI:
  10.1126/science.aab2277} {\bibfield  {journal} {\bibinfo  {journal}
  {Science}\ }\textbf {\bibinfo {volume} {350}},\ \bibinfo {pages} {1353}
  (\bibinfo {year} {2015})}\BibitemShut {NoStop}%
\bibitem [{\citenamefont {Tinkham}(2004)}]{Tinkham}%
  \BibitemOpen
  \bibfield  {author} {\bibinfo {author} {\bibfnamefont {M.}~\bibnamefont
  {Tinkham}},\ }\href@noop {} {\emph {\bibinfo {title} {Introduction to
  superconductivity}}}\ (\bibinfo  {publisher} {Courier Corporation},\ \bibinfo
  {year} {2004})\BibitemShut {NoStop}%
\bibitem [{\citenamefont {Buzea}\ and\ \citenamefont {Yamashita}(2001)}]{MgB2}%
  \BibitemOpen
  \bibfield  {author} {\bibinfo {author} {\bibfnamefont {C.}~\bibnamefont
  {Buzea}}\ and\ \bibinfo {author} {\bibfnamefont {T.}~\bibnamefont
  {Yamashita}},\ }\href {https://doi.org/10.1088/0953-2048/14/11/201}
  {\bibfield  {journal} {\bibinfo  {journal} {Superconductor Science and
  Technology}\ }\textbf {\bibinfo {volume} {14}},\ \bibinfo {pages} {R115}
  (\bibinfo {year} {2001})}\BibitemShut {NoStop}%
\bibitem [{\citenamefont {Eguchi}\ \emph {et~al.}(2013)\citenamefont {Eguchi},
  \citenamefont {Peets}, \citenamefont {Kriener}, \citenamefont {Yonezawa},
  \citenamefont {Bao}, \citenamefont {Harada}, \citenamefont {Inada},
  \citenamefont {Zheng},\ and\ \citenamefont {Maeno}}]{anis}%
  \BibitemOpen
  \bibfield  {author} {\bibinfo {author} {\bibfnamefont {G.}~\bibnamefont
  {Eguchi}}, \bibinfo {author} {\bibfnamefont {D.}~\bibnamefont {Peets}},
  \bibinfo {author} {\bibfnamefont {M.}~\bibnamefont {Kriener}}, \bibinfo
  {author} {\bibfnamefont {S.}~\bibnamefont {Yonezawa}}, \bibinfo {author}
  {\bibfnamefont {G.}~\bibnamefont {Bao}}, \bibinfo {author} {\bibfnamefont
  {S.}~\bibnamefont {Harada}}, \bibinfo {author} {\bibfnamefont
  {Y.}~\bibnamefont {Inada}}, \bibinfo {author} {\bibfnamefont {G.-q.}\
  \bibnamefont {Zheng}},\ and\ \bibinfo {author} {\bibfnamefont
  {Y.}~\bibnamefont {Maeno}},\ }\href
  {https://doi.org/https://doi.org/10.1103/PhysRevB.87.161203} {\bibfield
  {journal} {\bibinfo  {journal} {Physical Review B}\ }\textbf {\bibinfo
  {volume} {87}},\ \bibinfo {pages} {161203} (\bibinfo {year}
  {2013})}\BibitemShut {NoStop}%
\bibitem [{\citenamefont {Bardeen}\ \emph {et~al.}(1957)\citenamefont
  {Bardeen}, \citenamefont {Cooper},\ and\ \citenamefont {Schrieffer}}]{BCS}%
  \BibitemOpen
  \bibfield  {author} {\bibinfo {author} {\bibfnamefont {J.}~\bibnamefont
  {Bardeen}}, \bibinfo {author} {\bibfnamefont {L.~N.}\ \bibnamefont
  {Cooper}},\ and\ \bibinfo {author} {\bibfnamefont {J.~R.}\ \bibnamefont
  {Schrieffer}},\ }\href {https://doi.org/10.1103/PhysRev.108.1175} {\bibfield
  {journal} {\bibinfo  {journal} {Phys. Rev.}\ }\textbf {\bibinfo {volume}
  {108}},\ \bibinfo {pages} {1175} (\bibinfo {year} {1957})}\BibitemShut
  {NoStop}%
\bibitem [{\citenamefont {Johnston}(2013)}]{alpha}%
  \BibitemOpen
  \bibfield  {author} {\bibinfo {author} {\bibfnamefont {D.~C.}\ \bibnamefont
  {Johnston}},\ }\bibfield  {journal} {\bibinfo  {journal} {Supercond. Sci.
  Technol.}\ }\textbf {\bibinfo {volume} {26}},\ \href
  {https://doi.org/10.1088/0953-2048/26/11/115011}
  {10.1088/0953-2048/26/11/115011} (\bibinfo {year} {2013})\BibitemShut
  {NoStop}%
\bibitem [{\citenamefont {Maisuradze}\ \emph {et~al.}(2009)\citenamefont
  {Maisuradze}, \citenamefont {Khasanov}, \citenamefont {Shengelaya},\ and\
  \citenamefont {Keller}}]{secondmoment}%
  \BibitemOpen
  \bibfield  {author} {\bibinfo {author} {\bibfnamefont {A.}~\bibnamefont
  {Maisuradze}}, \bibinfo {author} {\bibfnamefont {R.}~\bibnamefont
  {Khasanov}}, \bibinfo {author} {\bibfnamefont {A.}~\bibnamefont
  {Shengelaya}},\ and\ \bibinfo {author} {\bibfnamefont {H.}~\bibnamefont
  {Keller}},\ }\href {https://doi.org/10.1088/0953-8984/21/7/075701} {\bibfield
   {journal} {\bibinfo  {journal} {Journal of Physics: Condensed Matter}\
  }\textbf {\bibinfo {volume} {21}},\ \bibinfo {pages} {075701} (\bibinfo
  {year} {2009})}\BibitemShut {NoStop}%
\bibitem [{\citenamefont {Brandt}\ and\ \citenamefont
  {Helmut}(2003)}]{penetrationdepth}%
  \BibitemOpen
  \bibfield  {author} {\bibinfo {author} {\bibnamefont {Brandt}}\ and\ \bibinfo
  {author} {\bibfnamefont {E.}~\bibnamefont {Helmut}},\ }\href
  {https://doi.org/10.1103/PhysRevB.68.054506} {\bibfield  {journal} {\bibinfo
  {journal} {Phys. Rev. B}\ }\textbf {\bibinfo {volume} {68}},\ \bibinfo
  {pages} {054506} (\bibinfo {year} {2003})}\BibitemShut {NoStop}%
\bibitem [{\citenamefont {Kabliman}\ \emph {et~al.}(2010)\citenamefont
  {Kabliman}, \citenamefont {Blaha},\ and\ \citenamefont {Schwarz}}]{PTS_DFT}%
  \BibitemOpen
  \bibfield  {author} {\bibinfo {author} {\bibfnamefont {E.}~\bibnamefont
  {Kabliman}}, \bibinfo {author} {\bibfnamefont {P.}~\bibnamefont {Blaha}},\
  and\ \bibinfo {author} {\bibfnamefont {K.}~\bibnamefont {Schwarz}},\ }\href
  {https://doi.org/10.1103/PhysRevB.82.125308} {\bibfield  {journal} {\bibinfo
  {journal} {Phys. Rev. B}\ }\textbf {\bibinfo {volume} {82}},\ \bibinfo
  {pages} {125308} (\bibinfo {year} {2010})}\BibitemShut {NoStop}%
\bibitem [{\citenamefont {Zhong}\ \emph {et~al.}(2023)\citenamefont {Zhong},
  \citenamefont {Zhang}, \citenamefont {Zhang}, \citenamefont {Bao},
  \citenamefont {Zhang}, \citenamefont {Xu}, \citenamefont {Luo}, \citenamefont
  {Rousuli}, \citenamefont {Yao}, \citenamefont {Denlinger}, \citenamefont
  {Huang}, \citenamefont {Wu}, \citenamefont {Xu}, \citenamefont {Duan},\ and\
  \citenamefont {Zhou}}]{PbSeNbSe2}%
  \BibitemOpen
  \bibfield  {author} {\bibinfo {author} {\bibfnamefont {H.}~\bibnamefont
  {Zhong}}, \bibinfo {author} {\bibfnamefont {H.}~\bibnamefont {Zhang}},
  \bibinfo {author} {\bibfnamefont {H.}~\bibnamefont {Zhang}}, \bibinfo
  {author} {\bibfnamefont {T.}~\bibnamefont {Bao}}, \bibinfo {author}
  {\bibfnamefont {K.}~\bibnamefont {Zhang}}, \bibinfo {author} {\bibfnamefont
  {S.}~\bibnamefont {Xu}}, \bibinfo {author} {\bibfnamefont {L.}~\bibnamefont
  {Luo}}, \bibinfo {author} {\bibfnamefont {A.}~\bibnamefont {Rousuli}},
  \bibinfo {author} {\bibfnamefont {W.}~\bibnamefont {Yao}}, \bibinfo {author}
  {\bibfnamefont {J.~D.}\ \bibnamefont {Denlinger}}, \bibinfo {author}
  {\bibfnamefont {Y.}~\bibnamefont {Huang}}, \bibinfo {author} {\bibfnamefont
  {Y.}~\bibnamefont {Wu}}, \bibinfo {author} {\bibfnamefont {Y.}~\bibnamefont
  {Xu}}, \bibinfo {author} {\bibfnamefont {W.}~\bibnamefont {Duan}},\ and\
  \bibinfo {author} {\bibfnamefont {S.}~\bibnamefont {Zhou}},\ }\href
  {https://doi.org/10.1103/PhysRevMaterials.7.L041801} {\bibfield  {journal}
  {\bibinfo  {journal} {Phys. Rev. Mater.}\ }\textbf {\bibinfo {volume} {7}},\
  \bibinfo {pages} {L041801} (\bibinfo {year} {2023})}\BibitemShut {NoStop}%
\bibitem [{\citenamefont {Yao}\ \emph {et~al.}(2018)\citenamefont {Yao},
  \citenamefont {Shen}, \citenamefont {Wen}, \citenamefont {Hua}, \citenamefont
  {Zhang}, \citenamefont {Wang}, \citenamefont {Niu}, \citenamefont {Chen},
  \citenamefont {Dudin}, \citenamefont {Lu}, \citenamefont {Zheng},
  \citenamefont {Chen}, \citenamefont {Wan},\ and\ \citenamefont
  {Feng}}]{PbSeTiSe2}%
  \BibitemOpen
  \bibfield  {author} {\bibinfo {author} {\bibfnamefont {Q.}~\bibnamefont
  {Yao}}, \bibinfo {author} {\bibfnamefont {D.~W.}\ \bibnamefont {Shen}},
  \bibinfo {author} {\bibfnamefont {C.~H.~P.}\ \bibnamefont {Wen}}, \bibinfo
  {author} {\bibfnamefont {C.~Q.}\ \bibnamefont {Hua}}, \bibinfo {author}
  {\bibfnamefont {L.~Q.}\ \bibnamefont {Zhang}}, \bibinfo {author}
  {\bibfnamefont {N.~Z.}\ \bibnamefont {Wang}}, \bibinfo {author}
  {\bibfnamefont {X.~H.}\ \bibnamefont {Niu}}, \bibinfo {author} {\bibfnamefont
  {Q.~Y.}\ \bibnamefont {Chen}}, \bibinfo {author} {\bibfnamefont
  {P.}~\bibnamefont {Dudin}}, \bibinfo {author} {\bibfnamefont {Y.~H.}\
  \bibnamefont {Lu}}, \bibinfo {author} {\bibfnamefont {Y.}~\bibnamefont
  {Zheng}}, \bibinfo {author} {\bibfnamefont {X.~H.}\ \bibnamefont {Chen}},
  \bibinfo {author} {\bibfnamefont {X.~G.}\ \bibnamefont {Wan}},\ and\ \bibinfo
  {author} {\bibfnamefont {D.~L.}\ \bibnamefont {Feng}},\ }\href
  {https://doi.org/10.1103/PhysRevLett.120.106401} {\bibfield  {journal}
  {\bibinfo  {journal} {Phys. Rev. Lett.}\ }\textbf {\bibinfo {volume} {120}},\
  \bibinfo {pages} {106401} (\bibinfo {year} {2018})}\BibitemShut {NoStop}%
\bibitem [{\citenamefont {Li}\ \emph {et~al.}(2024)\citenamefont {Li},
  \citenamefont {Lyu}, \citenamefont {Chen}, \citenamefont {Guan},
  \citenamefont {Yu}, \citenamefont {Zhao}, \citenamefont {Huang},
  \citenamefont {Zhou}, \citenamefont {Qiu}, \citenamefont {Fang},
  \citenamefont {Hashimoto}, \citenamefont {Lu}, \citenamefont {Song},
  \citenamefont {Loh}, \citenamefont {Zheng}, \citenamefont {Shen},
  \citenamefont {Novoselov},\ and\ \citenamefont {Lu}}]{SnSTaS2}%
  \BibitemOpen
  \bibfield  {author} {\bibinfo {author} {\bibfnamefont {Z.}~\bibnamefont
  {Li}}, \bibinfo {author} {\bibfnamefont {P.}~\bibnamefont {Lyu}}, \bibinfo
  {author} {\bibfnamefont {Z.}~\bibnamefont {Chen}}, \bibinfo {author}
  {\bibfnamefont {D.}~\bibnamefont {Guan}}, \bibinfo {author} {\bibfnamefont
  {S.}~\bibnamefont {Yu}}, \bibinfo {author} {\bibfnamefont {J.}~\bibnamefont
  {Zhao}}, \bibinfo {author} {\bibfnamefont {P.}~\bibnamefont {Huang}},
  \bibinfo {author} {\bibfnamefont {X.}~\bibnamefont {Zhou}}, \bibinfo {author}
  {\bibfnamefont {Z.}~\bibnamefont {Qiu}}, \bibinfo {author} {\bibfnamefont
  {H.}~\bibnamefont {Fang}}, \bibinfo {author} {\bibfnamefont {M.}~\bibnamefont
  {Hashimoto}}, \bibinfo {author} {\bibfnamefont {D.}~\bibnamefont {Lu}},
  \bibinfo {author} {\bibfnamefont {F.}~\bibnamefont {Song}}, \bibinfo {author}
  {\bibfnamefont {K.~P.}\ \bibnamefont {Loh}}, \bibinfo {author} {\bibfnamefont
  {Y.}~\bibnamefont {Zheng}}, \bibinfo {author} {\bibfnamefont {Z.-X.}\
  \bibnamefont {Shen}}, \bibinfo {author} {\bibfnamefont {K.~S.}\ \bibnamefont
  {Novoselov}},\ and\ \bibinfo {author} {\bibfnamefont {J.}~\bibnamefont
  {Lu}},\ }\href {https://doi.org/https://doi.org/10.1002/adma.202312341}
  {\bibfield  {journal} {\bibinfo  {journal} {Advanced Materials}\ }\textbf
  {\bibinfo {volume} {36}},\ \bibinfo {pages} {2312341} (\bibinfo {year}
  {2024})}\BibitemShut {NoStop}%
\bibitem [{\citenamefont {Leriche}\ \emph
  {et~al.}(2021{\natexlab{b}})\citenamefont {Leriche}, \citenamefont
  {Palacio-Morales}, \citenamefont {Campetella}, \citenamefont {Tresca},
  \citenamefont {Sasaki}, \citenamefont {Brun}, \citenamefont {Debontridder},
  \citenamefont {David}, \citenamefont {Arfaoui}, \citenamefont {Šofranko},
  \citenamefont {Samuely}, \citenamefont {Kremer}, \citenamefont {Monney},
  \citenamefont {Jaouen}, \citenamefont {Cario}, \citenamefont {Calandra},\
  and\ \citenamefont {Cren}}]{LaSeNbSe}%
  \BibitemOpen
  \bibfield  {author} {\bibinfo {author} {\bibfnamefont {R.~T.}\ \bibnamefont
  {Leriche}}, \bibinfo {author} {\bibfnamefont {A.}~\bibnamefont
  {Palacio-Morales}}, \bibinfo {author} {\bibfnamefont {M.}~\bibnamefont
  {Campetella}}, \bibinfo {author} {\bibfnamefont {C.}~\bibnamefont {Tresca}},
  \bibinfo {author} {\bibfnamefont {S.}~\bibnamefont {Sasaki}}, \bibinfo
  {author} {\bibfnamefont {C.}~\bibnamefont {Brun}}, \bibinfo {author}
  {\bibfnamefont {F.}~\bibnamefont {Debontridder}}, \bibinfo {author}
  {\bibfnamefont {P.}~\bibnamefont {David}}, \bibinfo {author} {\bibfnamefont
  {I.}~\bibnamefont {Arfaoui}}, \bibinfo {author} {\bibfnamefont
  {O.}~\bibnamefont {Šofranko}}, \bibinfo {author} {\bibfnamefont
  {T.}~\bibnamefont {Samuely}}, \bibinfo {author} {\bibfnamefont
  {G.}~\bibnamefont {Kremer}}, \bibinfo {author} {\bibfnamefont
  {C.}~\bibnamefont {Monney}}, \bibinfo {author} {\bibfnamefont
  {T.}~\bibnamefont {Jaouen}}, \bibinfo {author} {\bibfnamefont
  {L.}~\bibnamefont {Cario}}, \bibinfo {author} {\bibfnamefont
  {M.}~\bibnamefont {Calandra}},\ and\ \bibinfo {author} {\bibfnamefont
  {T.}~\bibnamefont {Cren}},\ }\href
  {https://doi.org/https://doi.org/10.1002/adfm.202007706} {\bibfield
  {journal} {\bibinfo  {journal} {Advanced Functional Materials}\ }\textbf
  {\bibinfo {volume} {31}},\ \bibinfo {pages} {2007706} (\bibinfo {year}
  {2021}{\natexlab{b}})}\BibitemShut {NoStop}%
\bibitem [{\citenamefont {Alla}\ \emph {et~al.}(2022)\citenamefont {Alla},
  \citenamefont {Gargee}, \citenamefont {Davide}, \citenamefont {Charlotte},
  \citenamefont {Marco}, \citenamefont {Nicola}, \citenamefont {Matthew},
  \citenamefont {Cephise}, \citenamefont {Martin},\ and\ \citenamefont
  {Philip}}]{BiSe_NbSe2_ARPES}%
  \BibitemOpen
  \bibfield  {author} {\bibinfo {author} {\bibfnamefont {C.}~\bibnamefont
  {Alla}}, \bibinfo {author} {\bibfnamefont {B.}~\bibnamefont {Gargee}},
  \bibinfo {author} {\bibfnamefont {C.}~\bibnamefont {Davide}}, \bibinfo
  {author} {\bibfnamefont {E.~S.}\ \bibnamefont {Charlotte}}, \bibinfo {author}
  {\bibfnamefont {B.}~\bibnamefont {Marco}}, \bibinfo {author} {\bibfnamefont
  {L.}~\bibnamefont {Nicola}}, \bibinfo {author} {\bibfnamefont
  {W.}~\bibnamefont {Matthew}}, \bibinfo {author} {\bibfnamefont
  {C.}~\bibnamefont {Cephise}}, \bibinfo {author} {\bibfnamefont
  {B.}~\bibnamefont {Martin}},\ and\ \bibinfo {author} {\bibfnamefont
  {H.}~\bibnamefont {Philip}},\ }\href@noop {} {\bibfield  {journal} {\bibinfo
  {journal} {Phys. Rev. Materials}\ }\textbf {\bibinfo {volume} {6}} (\bibinfo
  {year} {2022})}\BibitemShut {NoStop}%
\bibitem [{\citenamefont {Zhang}\ \emph {et~al.}(2014)\citenamefont {Zhang},
  \citenamefont {Liu}, \citenamefont {Luo}, \citenamefont {Freeman},\ and\
  \citenamefont {Zunger}}]{zhang2014hidden}%
  \BibitemOpen
  \bibfield  {author} {\bibinfo {author} {\bibfnamefont {X.}~\bibnamefont
  {Zhang}}, \bibinfo {author} {\bibfnamefont {Q.}~\bibnamefont {Liu}}, \bibinfo
  {author} {\bibfnamefont {J.-W.}\ \bibnamefont {Luo}}, \bibinfo {author}
  {\bibfnamefont {A.~J.}\ \bibnamefont {Freeman}},\ and\ \bibinfo {author}
  {\bibfnamefont {A.}~\bibnamefont {Zunger}},\ }\href
  {https://doi.org/https://doi.org/10.1038/nphys2933} {\bibfield  {journal}
  {\bibinfo  {journal} {Nature Physics}\ }\textbf {\bibinfo {volume} {10}},\
  \bibinfo {pages} {387} (\bibinfo {year} {2014})}\BibitemShut {NoStop}%
\bibitem [{\citenamefont {Bawden}\ \emph {et~al.}(2016)\citenamefont {Bawden},
  \citenamefont {Cooil}, \citenamefont {Mazzola}, \citenamefont {Riley},
  \citenamefont {Collins-McIntyre}, \citenamefont {Sunko}, \citenamefont
  {Hunvik}, \citenamefont {Leandersson}, \citenamefont {Polley}, \citenamefont
  {Balasubramanian} \emph {et~al.}}]{bawden2016spin}%
  \BibitemOpen
  \bibfield  {author} {\bibinfo {author} {\bibfnamefont {L.}~\bibnamefont
  {Bawden}}, \bibinfo {author} {\bibfnamefont {S.~P.}\ \bibnamefont {Cooil}},
  \bibinfo {author} {\bibfnamefont {F.}~\bibnamefont {Mazzola}}, \bibinfo
  {author} {\bibfnamefont {J.}~\bibnamefont {Riley}}, \bibinfo {author}
  {\bibfnamefont {L.}~\bibnamefont {Collins-McIntyre}}, \bibinfo {author}
  {\bibfnamefont {V.}~\bibnamefont {Sunko}}, \bibinfo {author} {\bibfnamefont
  {K.}~\bibnamefont {Hunvik}}, \bibinfo {author} {\bibfnamefont
  {M.}~\bibnamefont {Leandersson}}, \bibinfo {author} {\bibfnamefont
  {C.}~\bibnamefont {Polley}}, \bibinfo {author} {\bibfnamefont
  {T.}~\bibnamefont {Balasubramanian}}, \emph {et~al.},\ }\href
  {https://doi.org/https://doi.org/10.1038/ncomms11711} {\bibfield  {journal}
  {\bibinfo  {journal} {Nature communications}\ }\textbf {\bibinfo {volume}
  {7}},\ \bibinfo {pages} {11711} (\bibinfo {year} {2016})}\BibitemShut
  {NoStop}%
\bibitem [{\citenamefont {Simon}\ \emph {et~al.}(2024)\citenamefont {Simon},
  \citenamefont {Yerzhakov}, \citenamefont {Vakahi}, \citenamefont {Remennik},
  \citenamefont {Ruhman}, \citenamefont {Khodas}, \citenamefont {Millo},
  \citenamefont {Steinberg} \emph {et~al.}}]{TB1HTaS2}%
  \BibitemOpen
  \bibfield  {author} {\bibinfo {author} {\bibfnamefont {S.}~\bibnamefont
  {Simon}}, \bibinfo {author} {\bibfnamefont {H.}~\bibnamefont {Yerzhakov}},
  \bibinfo {author} {\bibfnamefont {A.}~\bibnamefont {Vakahi}}, \bibinfo
  {author} {\bibfnamefont {S.}~\bibnamefont {Remennik}}, \bibinfo {author}
  {\bibfnamefont {J.}~\bibnamefont {Ruhman}}, \bibinfo {author} {\bibfnamefont
  {M.}~\bibnamefont {Khodas}}, \bibinfo {author} {\bibfnamefont
  {O.}~\bibnamefont {Millo}}, \bibinfo {author} {\bibfnamefont
  {H.}~\bibnamefont {Steinberg}}, \emph {et~al.},\ }\href@noop {} {\bibfield
  {journal} {\bibinfo  {journal} {arXiv preprint arXiv:2405.12548}\ } (\bibinfo
  {year} {2024})}\BibitemShut {NoStop}%
\bibitem [{\citenamefont {Suter}\ and\ \citenamefont {Wojek}(2012)}]{musrfit}%
  \BibitemOpen
  \bibfield  {author} {\bibinfo {author} {\bibfnamefont {A.}~\bibnamefont
  {Suter}}\ and\ \bibinfo {author} {\bibfnamefont {B.}~\bibnamefont {Wojek}},\
  }\href {https://doi.org/https://doi.org/10.1016/j.phpro.2012.04.042}
  {\bibfield  {journal} {\bibinfo  {journal} {Physics Procedia}\ }\textbf
  {\bibinfo {volume} {30}},\ \bibinfo {pages} {69} (\bibinfo {year} {2012})},\
  \bibinfo {note} {12th International Conference on Muon Spin Rotation,
  Relaxation and Resonance (μSR2011)}\BibitemShut {NoStop}%
\bibitem [{\citenamefont {Fisk}\ and\ \citenamefont {Webb}(1976)}]{para2}%
  \BibitemOpen
  \bibfield  {author} {\bibinfo {author} {\bibfnamefont {Z.}~\bibnamefont
  {Fisk}}\ and\ \bibinfo {author} {\bibfnamefont {G.}~\bibnamefont {Webb}},\
  }\href {https://doi.org/https://doi.org/10.1103/PhysRevLett.36.1084}
  {\bibfield  {journal} {\bibinfo  {journal} {Physical Review Letters}\
  }\textbf {\bibinfo {volume} {36}},\ \bibinfo {pages} {1084} (\bibinfo {year}
  {1976})}\BibitemShut {NoStop}%
\bibitem [{\citenamefont {Gunnarsson}\ \emph {et~al.}(2003)\citenamefont
  {Gunnarsson}, \citenamefont {Calandra},\ and\ \citenamefont {Han}}]{para3}%
  \BibitemOpen
  \bibfield  {author} {\bibinfo {author} {\bibfnamefont {O.}~\bibnamefont
  {Gunnarsson}}, \bibinfo {author} {\bibfnamefont {M.}~\bibnamefont
  {Calandra}},\ and\ \bibinfo {author} {\bibfnamefont {J.}~\bibnamefont
  {Han}},\ }\href {https://doi.org/https://doi.org/10.1103/RevModPhys.75.1085}
  {\bibfield  {journal} {\bibinfo  {journal} {Reviews of Modern Physics}\
  }\textbf {\bibinfo {volume} {75}},\ \bibinfo {pages} {1085} (\bibinfo {year}
  {2003})}\BibitemShut {NoStop}%
\bibitem [{\citenamefont {Wiesmann}\ \emph {et~al.}(1977)\citenamefont
  {Wiesmann}, \citenamefont {Gurvitch}, \citenamefont {Lutz}, \citenamefont
  {Ghosh}, \citenamefont {Schwarz}, \citenamefont {Strongin}, \citenamefont
  {Allen},\ and\ \citenamefont {Halley}}]{para1}%
  \BibitemOpen
  \bibfield  {author} {\bibinfo {author} {\bibfnamefont {H.}~\bibnamefont
  {Wiesmann}}, \bibinfo {author} {\bibfnamefont {M.}~\bibnamefont {Gurvitch}},
  \bibinfo {author} {\bibfnamefont {H.}~\bibnamefont {Lutz}}, \bibinfo {author}
  {\bibfnamefont {A.}~\bibnamefont {Ghosh}}, \bibinfo {author} {\bibfnamefont
  {B.}~\bibnamefont {Schwarz}}, \bibinfo {author} {\bibfnamefont
  {M.}~\bibnamefont {Strongin}}, \bibinfo {author} {\bibfnamefont {P.~B.}\
  \bibnamefont {Allen}},\ and\ \bibinfo {author} {\bibfnamefont {J.~W.}\
  \bibnamefont {Halley}},\ }\href {https://doi.org/10.1103/PhysRevLett.38.782}
  {\bibfield  {journal} {\bibinfo  {journal} {Phys. Rev. Lett.}\ }\textbf
  {\bibinfo {volume} {38}},\ \bibinfo {pages} {782} (\bibinfo {year}
  {1977})}\BibitemShut {NoStop}%
\bibitem [{\citenamefont {Coffey}\ \emph {et~al.}(1985)\citenamefont {Coffey},
  \citenamefont {Levin},\ and\ \citenamefont {Muttalib}}]{hc21}%
  \BibitemOpen
  \bibfield  {author} {\bibinfo {author} {\bibfnamefont {L.}~\bibnamefont
  {Coffey}}, \bibinfo {author} {\bibfnamefont {K.}~\bibnamefont {Levin}},\ and\
  \bibinfo {author} {\bibfnamefont {K.}~\bibnamefont {Muttalib}},\ }\href
  {https://doi.org/https://doi.org/10.1103/PhysRevB.32.4382} {\bibfield
  {journal} {\bibinfo  {journal} {Physical Review B}\ }\textbf {\bibinfo
  {volume} {32}},\ \bibinfo {pages} {4382} (\bibinfo {year}
  {1985})}\BibitemShut {NoStop}%
\bibitem [{\citenamefont {Lebed}(1986)}]{hc22}%
  \BibitemOpen
  \bibfield  {author} {\bibinfo {author} {\bibfnamefont {A.}~\bibnamefont
  {Lebed}},\ }\href@noop {} {\bibfield  {journal} {\bibinfo  {journal} {ZhETF
  Pisma Redaktsiiu}\ }\textbf {\bibinfo {volume} {44}},\ \bibinfo {pages} {89}
  (\bibinfo {year} {1986})}\BibitemShut {NoStop}%
\bibitem [{\citenamefont {Klemm}\ \emph {et~al.}(1975)\citenamefont {Klemm},
  \citenamefont {Luther},\ and\ \citenamefont {Beasley}}]{hc23}%
  \BibitemOpen
  \bibfield  {author} {\bibinfo {author} {\bibfnamefont {R.~A.}\ \bibnamefont
  {Klemm}}, \bibinfo {author} {\bibfnamefont {A.}~\bibnamefont {Luther}},\ and\
  \bibinfo {author} {\bibfnamefont {M.}~\bibnamefont {Beasley}},\ }\href
  {https://doi.org/https://doi.org/10.1103/PhysRevB.12.877} {\bibfield
  {journal} {\bibinfo  {journal} {Physical Review B}\ }\textbf {\bibinfo
  {volume} {12}},\ \bibinfo {pages} {877} (\bibinfo {year} {1975})}\BibitemShut
  {NoStop}%
\bibitem [{\citenamefont {Gurevich}(2003)}]{hc24}%
  \BibitemOpen
  \bibfield  {author} {\bibinfo {author} {\bibfnamefont {A.}~\bibnamefont
  {Gurevich}},\ }\href {https://doi.org/10.1103/PhysRevB.67.184515} {\bibfield
  {journal} {\bibinfo  {journal} {Phys. Rev. B}\ }\textbf {\bibinfo {volume}
  {67}},\ \bibinfo {pages} {184515} (\bibinfo {year} {2003})}\BibitemShut
  {NoStop}%
\bibitem [{\citenamefont {Shulga}\ \emph {et~al.}(1998)\citenamefont {Shulga},
  \citenamefont {Drechsler}, \citenamefont {Fuchs}, \citenamefont {Muller},
  \citenamefont {Winzer}, \citenamefont {Heinecke},\ and\ \citenamefont
  {Krug}}]{YNBC}%
  \BibitemOpen
  \bibfield  {author} {\bibinfo {author} {\bibfnamefont {S.}~\bibnamefont
  {Shulga}}, \bibinfo {author} {\bibfnamefont {S.-L.}\ \bibnamefont
  {Drechsler}}, \bibinfo {author} {\bibfnamefont {G.}~\bibnamefont {Fuchs}},
  \bibinfo {author} {\bibfnamefont {K.-H.}\ \bibnamefont {Muller}}, \bibinfo
  {author} {\bibfnamefont {K.}~\bibnamefont {Winzer}}, \bibinfo {author}
  {\bibfnamefont {M.}~\bibnamefont {Heinecke}},\ and\ \bibinfo {author}
  {\bibfnamefont {K.}~\bibnamefont {Krug}},\ }\href
  {https://doi.org/https://doi.org/10.1103/PhysRevLett.80.1730} {\bibfield
  {journal} {\bibinfo  {journal} {Physical review letters}\ }\textbf {\bibinfo
  {volume} {80}},\ \bibinfo {pages} {1730} (\bibinfo {year}
  {1998})}\BibitemShut {NoStop}%
\bibitem [{\citenamefont {Suderow}\ \emph {et~al.}(2004)\citenamefont
  {Suderow}, \citenamefont {Tissen}, \citenamefont {Brison}, \citenamefont
  {Martinez}, \citenamefont {Vieira}, \citenamefont {Lejay}, \citenamefont
  {Lee},\ and\ \citenamefont {Tajima}}]{YNBC2}%
  \BibitemOpen
  \bibfield  {author} {\bibinfo {author} {\bibfnamefont {H.}~\bibnamefont
  {Suderow}}, \bibinfo {author} {\bibfnamefont {V.}~\bibnamefont {Tissen}},
  \bibinfo {author} {\bibfnamefont {J.}~\bibnamefont {Brison}}, \bibinfo
  {author} {\bibfnamefont {J.}~\bibnamefont {Martinez}}, \bibinfo {author}
  {\bibfnamefont {S.}~\bibnamefont {Vieira}}, \bibinfo {author} {\bibfnamefont
  {P.}~\bibnamefont {Lejay}}, \bibinfo {author} {\bibfnamefont
  {S.}~\bibnamefont {Lee}},\ and\ \bibinfo {author} {\bibfnamefont
  {S.}~\bibnamefont {Tajima}},\ }\href
  {https://doi.org/https://doi.org/10.1103/PhysRevB.70.134518} {\bibfield
  {journal} {\bibinfo  {journal} {Physical Review B}\ }\textbf {\bibinfo
  {volume} {70}},\ \bibinfo {pages} {134518} (\bibinfo {year}
  {2004})}\BibitemShut {NoStop}%
\bibitem [{\citenamefont {Suderow}\ \emph {et~al.}(2005)\citenamefont
  {Suderow}, \citenamefont {Tissen}, \citenamefont {Brison}, \citenamefont
  {Mart\'{\i}nez},\ and\ \citenamefont {Vieira}}]{NbSe2}%
  \BibitemOpen
  \bibfield  {author} {\bibinfo {author} {\bibfnamefont {H.}~\bibnamefont
  {Suderow}}, \bibinfo {author} {\bibfnamefont {V.~G.}\ \bibnamefont {Tissen}},
  \bibinfo {author} {\bibfnamefont {J.~P.}\ \bibnamefont {Brison}}, \bibinfo
  {author} {\bibfnamefont {J.~L.}\ \bibnamefont {Mart\'{\i}nez}},\ and\
  \bibinfo {author} {\bibfnamefont {S.}~\bibnamefont {Vieira}},\ }\href
  {https://doi.org/10.1103/PhysRevLett.95.117006} {\bibfield  {journal}
  {\bibinfo  {journal} {Phys. Rev. Lett.}\ }\textbf {\bibinfo {volume} {95}},\
  \bibinfo {pages} {117006} (\bibinfo {year} {2005})}\BibitemShut {NoStop}%
\bibitem [{\citenamefont {Li}\ \emph {et~al.}(2019)\citenamefont {Li},
  \citenamefont {Tabis}, \citenamefont {Tang}, \citenamefont {Yu},
  \citenamefont {Jaroszynski}, \citenamefont {Barišić},\ and\ \citenamefont
  {Greven}}]{twogapeqn}%
  \BibitemOpen
  \bibfield  {author} {\bibinfo {author} {\bibfnamefont {Y.}~\bibnamefont
  {Li}}, \bibinfo {author} {\bibfnamefont {W.}~\bibnamefont {Tabis}}, \bibinfo
  {author} {\bibfnamefont {Y.}~\bibnamefont {Tang}}, \bibinfo {author}
  {\bibfnamefont {G.}~\bibnamefont {Yu}}, \bibinfo {author} {\bibfnamefont
  {J.}~\bibnamefont {Jaroszynski}}, \bibinfo {author} {\bibfnamefont
  {N.}~\bibnamefont {Barišić}},\ and\ \bibinfo {author} {\bibfnamefont
  {M.}~\bibnamefont {Greven}},\ }\href {https://doi.org/10.1126/sciadv.aap7349}
  {\bibfield  {journal} {\bibinfo  {journal} {Science Advances}\ }\textbf
  {\bibinfo {volume} {5}},\ \bibinfo {pages} {eaap7349} (\bibinfo {year}
  {2019})},\ \Eprint
  {https://arxiv.org/abs/https://www.science.org/doi/pdf/10.1126/sciadv.aap7349}
  {https://www.science.org/doi/pdf/10.1126/sciadv.aap7349} \BibitemShut
  {NoStop}%
\end{thebibliography}%

\appendix

\section{Transport}
\label{Appendix:res}
\setcounter{table}{0}
\renewcommand{\thetable}{\thesection.\arabic{table}}

The resistivity data, covering a temperature range from \(10 \mathrm{K}\) to \(400 \mathrm{K}\), exhibit a gradual decrease in the slope of \(\rho(T)\), deviating from linear bahvior, typically attributed to the shortening of the mean free path to a few inter-atomic spacings. This behavior, which deviates from the linear scattering perturbation model, is primarily due to electron-phonon interactions dominating at high temperatures. As a result, resistivity increases less rapidly with temperature \cite{para2,para3}, a behavior that can be effectively modeled using the parallel resistor model \cite{para1}.

The generalized resistivity model fit for \( T > T _c \), as shown in  \cref{fig:PTS_Res}a, considers both temperature-independent residual resistivity arising from impurities and disorder (\(\rho_{i,0}\)) and a temperature-dependent term formulated according to the generalized Bloch-Grüneisen expression:
\begin{equation}
\rho_{i,L}(T) = C\left(\frac{T}{\Theta_{D}}\right)^5 \int_{0}^{\Theta_{D}/T} \frac{x^5}{(e^x-1)(1-e^{-x})}dx
\end{equation}
where \(\Theta_{D}\) is the Debye temperature derived from resistivity measurements, and \(C\) is a material-specific constant. At high temperatures, the resistivity curve, considering saturation, is explained by:
\begin{equation}
\rho(T) = \left[\frac{1}{\rho_s} + \frac{1}{\rho_{i,0} + \rho_{i,L}(T)}\right]^{-1}
\label{eq:rhoT}
\end{equation}
where \(\rho_s\) represents the temperature-independent saturation resistivity value reached at high temperatures. The fit yields a Debye temperature of \(\Theta_{D} = 178(3) \K\), \(C = 2.23 (4) \; \mathrm{m\Omega \cdot cm}\), residual resistivity \(\rho_{i,0} = 91(1) \;\mathrm{\mu\Omega \cdot cm}\), and saturation resistivity \(\rho_s = 848(4) \;\mathrm{\mu\Omega \cdot cm}\).\\

The temperature dependence of the upper critical field \(H_{c2}(T)\) in both the out-of-plane and in-plane directions exhibited an atypical upturn at higher magnetic fields, diverging from the behavior predicted by conventional models such as the Ginzburg-Landau (G-L) or Werthamer-Helfand-Hohenberg (WHH) models \cite{WHH}. This upturn may result from various reasons, including localization effects \cite{hc21}, magnetic field-induced twisting of electron orbits \cite{hc22}, dimensional crossover \cite{hc23} and multigap superconductivity \cite{hc24}. A similar upturn was observed in  \(\mathrm{MgB_2}\) \cite{MgB2}, \(\mathrm{YNi_2 B_2 C}\) \cite{YNBC}, \(\mathrm{LuNi_2 B_2 C}\) \cite{YNBC2}, and \(\mathrm{2H-NbSe_2}\) \cite{NbSe2} and was attributed to two-gap superconductivity.

According to two-band model, \(H_{c2}(T)\) can be expressed as  \cite{twogapeqn}:
 \begin{multline}
\ln\left(\frac{T}{T_c}\right) = \left[U(s) + U\left(\eta\right) + \frac{\lambda_{0}}{w}\right] + \\
\sqrt{\left(\frac{1}{4} \left[U(s) - U\left(\eta s\right) - \frac{\lambda_{-}}{w}\right]^2 + \frac{\lambda _{12}\lambda _{21}}{w^2}\right)}
\label{eq:whha}
\end{multline}
Here, \( \lambda_{-} = \lambda_{11} - \lambda_{22} \), \( \lambda_{0} = \sqrt{\lambda_{-}^2 + 4\lambda_{12}\lambda_{21}}\), and \( w = \lambda_{11}\lambda_{22} - \lambda_{21}\lambda_{12} \). The variables \( \lambda_{11}, \lambda_{22}, \lambda_{12}\), and \(\lambda_{21} \) represent the matrix elements of the BCS coupling constants.
The electron and hole diffusivity are denoted as \(D_{1}\) and \(D_{2}\), is accounted by $\eta = (D_{2}/D_{1})$. The function \(U(s)\) is defined using the digamma function \(\psi(s)\) as \(U(s) = \psi\left(s + \frac{1}{2}\right) - \psi\left(\frac{1}{2}\right)\). The upper critical field is related to these parameters as follows:
 \begin{equation}
H_{c2}   = \frac{ 2\phi_{0}Ts}{D_{e} }
\label{eq:whhb}
\end{equation}
where \( \phi_{0} \) is the magnetic flux quantum. 

During the fitting, we observed intraband coupling terms $\lambda_{11}$ and $\lambda_{22}$ converges to close values, similarly for intraband coupling $\lambda_{12}$ and $\lambda_{21}$. For the final fitting, we have kept $\lambda_{22}$ = $\lambda_{11}$, $\lambda_{12}$ = $\lambda_{21}$, and $T_c$ = 3.08 K was fixed. The resulting fitting parameters are given in \cref{table}. The small value of $\eta$ indicates that the scattering in one of the bands is larger compared to the other. Also, the intraband coupling is found to be stronger than the interband coupling, indicating the two bands behave almost independently, leading to two distinct gaps. The fitting done using this model gave the upper critical field values as \(0.57 \T\)  and \(6.7 \T\)  at \(0 \K\). \\

\begin{table}[h!]
\centering
\begin{tabular}{ c  c  c  c  c  } 
\hline

type & $D_{1}$ & $\eta$ & $\lambda_{11}$ & $\lambda_{12}$ \\
\hline
\hline
$H_{c2} \parallel c$ & 11.47 & 0.237 & 1.432 & 0.274\\
\hline
$H_{c2} \parallel ab$ & 1.42 & 0.157 & 1.41 & 0.36\\
\hline
\end{tabular}
\caption{Fitting parameters from two-gap model for $H_{c2}$}
\label{table}
\end{table}

\section{Band dispersion along k$_z$}
\label{Appendix:KZ}
In this appendix, we detail the method used to determine \(k_\bot\) values by employing the free-electron state approximation. In this model, an electron is excited by a photon and assumed to adhere to the dispersion relation \(E \propto k^2_\bot\). When the electron transitions into the vacuum, it loses momentum as it overcomes the inner potential step barrier \(V_0\), set here at \(9 \eV\), which is dependent on the material.

The formula for \(k_\bot\) inside the crystal is expressed as:

\[
k_\bot = \frac{1}{\hbar} \sqrt{2m_e \left(E_k \cos^2 \vartheta + V_0\right)}
\]

Here, \(E_k\) represents the kinetic energy of the electron, and \(\vartheta\) is the angle of electron ejection relative to the normal surface.

The value of \(k_z\) is determined by averaging across all values of \(E_k\) and \(\vartheta\). This methodological approach results in an approximate error margin of approximately 5\%.

\section{PTS core level}
\label{Appendix:core}
\setcounter{figure}{0}
\renewcommand{\thefigure}{\thesection.\arabic{figure}}
We measured the core-level spectra of our sample over an energy range of 26 eV to 18 eV below the Fermi level. The spectra show no significant \(k_\parallel\)-dependence and the profile is presented in \cref{fig:corelevel}.

\begin{figure}[ht!]
\includegraphics[width=0.95\columnwidth]{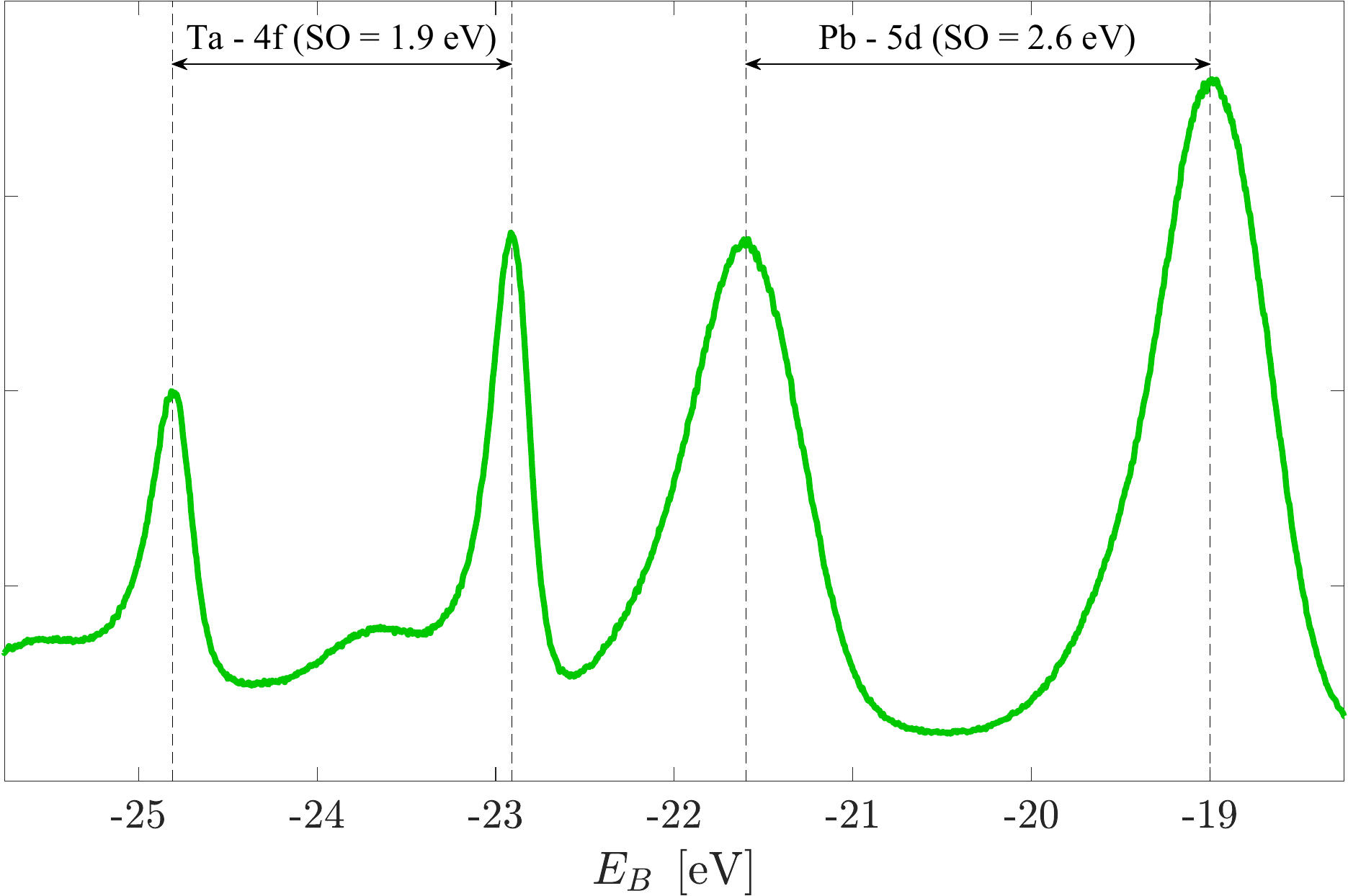}
\caption{Core-level spectra at the measurement point.}
\label{fig:corelevel}
\end{figure}

In the spectra, we observe four distinct peaks. Two of these peaks correspond to the \(4f_{5/2}\) and \(4f_{7/2}\) orbitals of Ta atoms, separated by the expected SO splitting of \(1.9 \, \mathrm{eV}\). The other two peaks arise from the Pb atoms, with a spin-orbit splitting \( 2.6 \, \mathrm{eV}\), corresponding to the \(5d_{5/2}\) and \(5d_{3/2}\) orbitals. This observation confirms that we are measuring contributions from both the PbS and TaS\(_2\) layers, indicating that the signal originates from both components of the heterostructure.

\end{document}